\documentclass[12pt]{iopart}

\usepackage[us,24hr]{datetime} 

\usepackage{lineno}

\usepackage[pdftex]{graphics,color}
\renewcommand{\textcolor}[2]{#2}
\renewcommand{\color}[1]{ }
\renewcommand{\normalcolor}{ }

\newcommand{\citep}{\cite}
\newcommand{\citet}{\cite}

\begin{document}

\title[The rotation problem
\hfill \today, \currenttime \hfill
]{An approximate application of quantum gravity to the rotation problem}

\author{
R Michael Jones
}

\address{
CIRES, University of Colorado Boulder \\ Boulder, CO 80309-0216
}
\ead{
Michael.Jones@Colorado.edu
}
\vspace{10pt} \begin{indented} \item[]\today, \currenttime \end{indented}

\begin{abstract}
\textcolor {blue}{Arbitrary initial conditions allow solutions of Einstein's field equations for General Relativity to have arbitrarily large relative rotation of matter and inertial frames.  The ``Rotation Problem'' is to explain why the measured relative rotation rate is so small.  As it turns out, nearly any reasonable theory of quantum gravity can solve the rotation problem by phase interference.}  Even as early as \textcolor {red}{about a quarter of a second after the initial simgularity, quantum cosmology would limit the cosmologies that contribute significantly to a path integral calculation to have relative rms rotation rates less than about} \textcolor {blue}{$10^{-51}$ radians per year.}  Those calculations are based on using 50 e-foldings during inflation.  For 55 or 60 e-foldings, the cosmologies contributing significantly to the path integral would have even smaller relative rotation rates.  In addition, although inflation dominates the calculation, even if there had been no inflation, the cosmologies contributing significantly to the path integral would have relative rotation rates less than about \textcolor {red}{$10^{-32}$ radians per year at about a quarter of a second after the initial singularity.}  These calculations are insensitive to the details of the theory of quantum gravity because the main factor depends only on the size of the visible universe, the Planck time, the free-space speed of light, the Hubble parameter, and the number of e-foldings during inflation.  These calculations use the Einstein-Hilbert action in quantum gravity, \textcolor {blue}{including} large-scale relative rotation of inertial frames and the matter distribution, in which each ``path'' is a cosmology with a different rms relative rotation rate.  The calculations include inflation for 50, 55, and 60 e-foldings, and for values of the dependence of relative rotation rate on cosmological scale factor $a$ as $a^{-m}$ for various values of $m$.  The calculation shows that the action is an extremum at zero rms relative rotation rate.  

\end{abstract}

%
\vspace{2pc}

%
\vspace{1pc}
\submitto{\CQG}
Accepted for publication 6 March 2024
%
%
%

\bibliographystyle{unsrt}

\section{\label{Rotate.intro}      Introduction}

\textcolor {red}{Arbitrary initial conditions allow solutions of Einstein's field equations for General Relativity to have large-scale relative rotation of matter and inertial frames e.g. \citep{Kramer-Stephani-MacCallum-Herlt:1980,Stephani-Kramer-MacCallum-Hoenselaers-Herlt:2003,EllisMacCallum69,1996gpst.conf..421K,Chechin:2013}.
The ``Rotation Problem'' \cite{Ellis-Olive:1983} is to explain why 
the measured rms} \textcolor {red}{(root mean square)} relative rotation rate is so small e.g. \citep{Barrow-Juszkiewicz-Sonoda:1985,Bayin-Cooperstock:1980,Collins-Hawking:1973b,Collins-Hawking:1973a,Ellis:1971,Ellis:2006,Ellis:2009,Ellis-Wainwright:1997,Fennelly:1976,Hawking69,Jaffe.et.al:2005,Jaffe.et.al:2006,Raine-Thomas:1982,PhysRevLett.117.131302,Su-Chu:2009,Wolfe:1970}.
The most restrictive of these measurements is less than about $10^{-20}$ radians per year.  Although most of these measurements are based on the cosmic microwave background radiation, the measurements by Wolfe \citep{Wolfe:1970} are based on X-ray measurements.  It is difficult to explain the absence of relative rotation in our universe classically without assuming very finely tuned initial conditions. 

\color {red}
Hawking and Luttrell \citep{HAWKING:LUTTRELL:1984} show that using a path integral over compact metrics without boundary would lead to a quantum state that is very nearly isotropic at the present time.  That is, they \textcolor {blue}{assume particular initial conditions rather than considering arbitrary} initial conditions.  Amsterdamski \citep{Amsterdamski:PhysRevD.31.3073} also uses a path integral over compact metrics, \textcolor {blue}{assuming particular} initial conditions to get small anisotropy at the present time.

Moss and Wright \citep{Moss:Wright:PhysRevD.29.1067} show that assuming a particular form for inflation and the Higgs field leads to a small amount of anisotropy at the present time.  Wright and Moss \citep{WRIGHT:MOSS:1985:115} numerically calculate the wavefunction for a Bianchi IX universe.  They show that the isotropy in the cosmic microwave background bears the imprint of the classical or quantum origin of the universe.  

Ellis and Olive \citep{Ellis-Olive:1983} use inflation to explain the lack of observed relative rotation.  

As pointed out by Hawking and Luttrell \citep{HAWKING:LUTTRELL:1984}, however, ``one could always make a cosmological model with an arbitrary amount of anisotropy at the present time and simply run the model back in time to find out what initial state it must have started from.  A similar objection applies to the inflationary model of the universe''
\normalcolor

A path-integral calculation in quantum cosmology that includes initial conditions \cite{Jones:rotation:problem:2020} showed that the action had an extremum at zero rms relative rotation rate and that the maximum relative rotation rate of cosmologies that would contribute to the path integral would be about $10^{-73}$ radians per year at the present time neglecting inflation, with inflation limiting the maximum relative rotation rate even more.  Because that calculation included only classical cosmologies, the action also had an extremum at zero initial relative rotation rate.

A question not answered by \cite{Jones:rotation:problem:2020}, however, is how the maximum rms relative rotation rate would vary with cosmological time.  It is the purpose here to address that question by using a path integral in which the action would be calculated by integrating the Lagrangian from the beginning of inflation to an arbitrary specified time. 

Section \ref{Rotate.QuantumGravity} argues that even if we do not have a generally accepted theory of quantum gravity, there are cases, such as the present one, in which we know enough about quantum gravity to make order-of-magnitude calculations. 
Section \ref{Rotate.path} reviews path integrals in quantum cosmology.
Section \ref{Rotate.action.approx} gives the approximate formula for the action for small vorticity.
Section \ref{Rotate.Saddlepoint} points out that the path integral has a saddlepoint at zero rms relative rotation, and gives the saddlepoint approximation.
Section \ref{Rotate.results} presents the results of the calculations of the maximum large-scale relative rotation as a function of global time.
Section \ref{Rotate.Disc} discusses the results.

\ref{Rotate.simple} gives the Friedmann equation for the background cosmology.
\ref{Rotate.main} gives the formula for the action.
\ref{Rotate.Lagrangian.App} gives the approximate {Lagrangian} for small vorticity.
\ref{Rotate.acceleration} calculates the relative rotation angle.  
\ref{Rotate.ell} calculates the deviation of the local scale factor $\ell$ along flow lines from the cosmological scale factor $a$. 
\ref{Rotate.Friedmann} gives the approximate generalized Friedmann equation for small rotation.
\ref{Rotate.correct} and \ref{Rotate.noinflation} give the formulas for the effect of small vorticity including and neglecting the effect of inflation, respectively.

The free-space speed of light $c$ and Newton's gravitational constant $G$ are taken as one except when converting to ordinary units.

\section{\label{Rotate.QuantumGravity}      Quantum Gravity}

Uniting General Relativity and quantum theory is one of the most important problems in theoretical physics.  Although this project is usually referred to as developing a theory of quantum gravity (that is, quantizing the gravitational field), that terminology may be misleading in that it will probably be necessary to revise both General Relativity and quantum theory to develop a quantum theory of gravity.  

The problem is very difficult e.g. \citep{isham1975quantum,isham1981quantum,smolin:2001:three,DeWitt:1967a,DeWitt:1967b,DeWitt:1967c,Wheeler68,Giulini2009,Kiefer2009,Kiefer2013},
 and it is not likely to be solved soon.  For example, a path-integral representation of quantum gravity requires knowing the action and the measure and it is necessary to handle the problem of diffeomorphisms.  

In the meantime, however, it is possible to make some calculations that are somewhat insensitive to the details of the correct theory of quantum gravity.  That allows us to make quantum gravity calculations in some cases of interest where only order-of-magnitude calculations are needed.  One example of such a calculation is presented here.  Using a correct theory of quantum gravity would probably not change the results by more than a factor of a few powers of 10.

\section[      Review of path integrals]{\label{Rotate.path}       Review of path integrals in quantum cosmology}

Although the term ``path integral'' did not exist at the time of Fermat, the concept did exist then.  Fermat's principle (which is based on Huygens principle) asks ``For a wave source at location A, which path (or paths) between A and B contribute most to the arrival of the wave at location B?''  The answer is those paths for which the action is an extremum for small variations of the path.  Because wave phase is proportional to the action, phase is also an extremum for the Fermat paths.  In addition, there are paths near the Fermat path whose phase differs from that of the Fermat path by less than $90^{\circ}$.  For these paths within the center of the first Fresnel zone, there is constructive interference.  

This same concept is generalized in quantum cosmology, in which we consider all of the ``paths'' connecting some initial 3-geometry with some final 3-geometry.  In this case, each ``path'' is a 4-geometry.  

Following Hartle and Hawking \citep{Hartle-Hawking83}, a path-integral calculation gives
\begin{equation}
	\fl
	\psi_f(^{(3)}\mathcal{G}_f, \phi_f) 
	= \int
	\int
	\exp{(i I[^{(4)}\mathcal{G},\phi]/\hbar)}
	\, \mathcal{D} {^{(4)}\mathcal{G}} \, \mathcal{D} \phi \, 
	\psi_i(^{(3)}\mathcal{G}_i, \phi_i) 
	\, \mathcal{D} {^{(3)}\mathcal{G}_i} \, \mathcal{D} \phi_i \, 
	\mbox{ , }
	\label{Rotate.relative-acceleration0e}
\end{equation}
where $\psi_i(^{(3)}\mathcal{G}_i, \phi_i)$ is the amplitude that the 3-geometry was $^{(3)}\mathcal{G}_i$ on some initial space-like hypersurface and the matter fields on that 3-geometry were $\phi_i$,  $\psi_f(^{(3)}\mathcal{G}_f, \phi_f)$ is the amplitude that the 3-geometry is $^{(3)}\mathcal{G}_f$ on some final space-like hypersurface and that the matter fields on that 3-geometry are $\phi_f$, $\exp{(i I[^{(4)}\mathcal{G},\phi]/\hbar)}$ is the contribution of the 4-geometry $^{(4)}\mathcal{G}$ and matter field $\phi$ on that 4-geometry to the path integral, where $I[^{(4)}\mathcal{G},\phi]$ is the action.  A correct theory of quantum gravity would be necessary to specify the measures $\mathcal{D} {^{(4)}\mathcal{G}}$ and $\mathcal{D} \phi$ and the action $I$.  \textcolor {red}{Although accurate calculations of $\psi_f(^{(3)}\mathcal{G}_f, \phi_f)$ in (\ref{Rotate.relative-acceleration0e}) would require knowing the measures $\mathcal{D} {^{(4)}\mathcal{G}}$ and $\mathcal{D} \phi$ \citep{Halliwell:PhysRevD.38.2468:1988,PhysRevD.96.106005} and the action $I$, it is possible in some simple cases to make order-of-magnitude estimates in a path integral without knowing details of the theory of quantum gravity.} 

Although in (\ref{Rotate.relative-acceleration0e}), the integration is over all possible 4-geometries, not just classical 4-geometries, the main contribution to the integral (in most cases) comes from classical 4-geometries, e.g. \citep{Kiefer:1991,Halliwell-Hartle:1990}.  That allows us to  restrict (\ref{Rotate.relative-acceleration0e}) to be an integration over classical 4-geometries that are solutions of Einstein's field equations.  
\textcolor {red}{Such classical 4-geometries in this case can be characterized by the initial rms vorticity, the final rms vorticity, or any rms vorticity in between.  Because of that, although the path integral is in principle an integral over the initial rms vorticity, calculating the path integral in this case as an integral over the final rms vorticity is equivalent to an integral over the initial rms vorticity.}

The condition that there are not finely tuned initial conditions is equivalent to $\psi_i(^{(3)}\mathcal{G}_i, \phi_i)$ being a broad wave function.  That allows us to neglect the effect of that initial wave function on the integration in the path integral in (\ref{Rotate.relative-acceleration0e}).  In addition, we consider 4-geometries characterized by a parameter ${\langle{\omega}_f\rangle}$ which we take to be the rms relative rotation rate on the space-like hypersurface at $t=t_f$.  Thus, we can rewrite (\ref{Rotate.relative-acceleration0e}) for our purposes as
\begin{equation}
	\psi_f(^{(3)}\mathcal{G}_f, \phi_f) 
	=  
	\int_{-\infty}^{\infty} 
	A(\langle{\omega}_f\rangle) 
	e^{i I(\langle{\omega}_f\rangle)/\hbar}
	\, \mathrm{d}{\langle{\omega}_f\rangle} 
	\mbox{ , }
	\label{Rotate.relative-acceleration0c}
\end{equation}
where $A(\langle{\omega}_f\rangle)$  is a slowly varying function of $\langle{\omega}_f\rangle$ \textcolor {red}{that includes the measure}, $I(\langle{\omega}_f\rangle)$ \textcolor {red}{given by (\ref{Rotate.6.1.c.1a})} is the action, and since for classical spacetimes, the rotation rate is a known function of cosmological time, we can consider the action $I$ to depend on the rms rotation rate at any cosmological time we choose, say $t_f$, which we designate as $\langle{\omega}_f\rangle$, where $\langle{\omega}_f\rangle^2\equiv\overline{{\omega}_f^2}$, and the average is a spatial average over the volume.  There is no assumption that the relative rotation rate is either homogeneous or isotropic.  The action depends only on the rms value of the relative rotation rate for small ${\langle{\omega}_f\rangle}$.

\section{\label{Rotate.action.approx}      Approximate action}
\textcolor {blue}{The derivation of the action including relative rotation is straightforward but very lengthy.  To include the derivation in detail would increase the length of the manuscript by a great deal.  On the other hand, to not include a description for how to do the derivation would not allow a diligent reader the possibility to verify the final results.  Therefore, the appendices include all of the formulas needed for the derivation.  Here is how to use those formulas to calculate the action.}  

\textcolor {blue}{We start with the action in (\ref{Rotate.6.1.c.1a}).  Then we use (\ref{Rotate.sqrt.g}) for $(-g^{(4)})^{1/2}$, (\ref{Rotate.L.sec.order}) for ${L}$, (\ref{Rotate.adot}) for $1/\dot{a}$, and (\ref{Rotate.volume}) for $V(a)$.  This brings in many new variables.  At this point, we use (\ref{Rotate.f.g}) for $f_g(a)$, (\ref{Rotate.F.of.a})   for $F(a)$ and (\ref{Rotate.Friedmann.f.H}) for $f_H(a)$.  This brings in more variables.  At this point, we use  (\ref{Rotate.H}) for $H(a)$ except during inflation,  (\ref{Rotate.H.Inflation}) for $H(a)$ during inflation,  (\ref{Rotate.p.w.rho}) for $p$,  (\ref{Rotate.rho.of.r}) for $\rho$, (\ref{Rotate.F.sub.L.of.a}) for $f_L(a)$,  (\ref{Rotate.f.ell}) for $f_\ell(a)$,  (\ref{Rotate.f.theta}) for $f_\theta(a)$, (\ref{Rotate.F.sub.a}) for $F_{\omega}(a)$, and  (\ref{Rotate.f.a}) for $f_{\omega}(a)$.  This gives the formula for the action to lowest order in the rms rotation rate  $\langle{\omega}_f\rangle$ as}
\begin{eqnarray}
	&
	I(\langle{\omega}_f\rangle) \approx 
	I_0 + \hbar \left(\frac{\langle{\omega}_f\rangle}{{\omega}_m}\right)^2  
\left[D_I(a_f) + \frac{\langle\omega_f\rangle^2+\sigma_{\omega}^2/\langle\omega_f\rangle^2}{H_f^2} 
D_{II}(a_f)\right] 
\nonumber \\ 
&
\approx 
	I_0 + \hbar \left(\frac{\langle{\omega}_f\rangle}{{\omega}_m}\right)^2  
	D_I(a_f) 
	\mbox{ , }
	\label{Rotate.small.average.omega}
\end{eqnarray}
where $I_0$ is the action for the standard cosmological model (FLRW cosmology \citep{Friedmann:1922,LeMaitre:1927,Robertson:1935ApJ,Robertson:1936ApJ-a,Robertson:1936ApJ-b,Walker:1937}),
 $\langle{\omega}_f\rangle$ is the rms relative rotation rate at the final global time $t_f$, 
\begin{equation}
	{\omega}_m
	= 
	\left( \frac{ \hbar 
		H_f
	}{\textcolor {red}{a_f^3} r_f^3 } \right)^{1/2}
	\to  {T^*}\sqrt{\frac{H_fc^3}{\textcolor {red}{a_f^3}r_f^3}}
	=  {T^*}\sqrt{\frac{H_fc^3}{\textcolor {red}{a_f^3}r_0^3 a_f^3}}
	=  \frac{{T^*} {H_f^{5/4}}}{(H_f a_f^2)^{3/4}} \left(\frac{c}{\textcolor {blue}{a_f}r_0}\right)^{3/2}
	\mbox{ , }
	\label{Rotate.omegaSub.m.2}
\end{equation}
$H_f$ is the Hubble parameter at the time $t_f$, $\sigma_{\omega}^2$ is the variance of $\omega_f^2$, $r_f$ is the radius of the visible universe at the time $t_f$, $a_f$ is the value of the cosmological scale factor at the time $t_f$, $r_0 \approx 46.5 \times 10^9$ light years is the present radius of the visible universe, $T^*\approx 1.7\times 10^{-51}$ yr is the Planck time, 
\begin{eqnarray}
	&&
	D_I(a_f) 
	= \frac{ H_0^2 \Omega_\Lambda}{H_fa_f^3} \left(\frac{2\pi}{3} \alpha_3 - \frac{ 
		\alpha_2}{12}  \right) \textcolor {red}{\frac{1}{a_f^3}} \int_{a_i}^{a_f}  \frac{ \textcolor {red}{ a^5}    }{  H(a)} [f_H(a) \textcolor {red}{+ f_g(a)}]
	\, \mathrm{d} \, a
	\nonumber \\ &&
	+ \frac{2\pi  H_f}{9 a_f^3}  \textcolor {red}{\frac{1}{a_f^3}} \int_{a_i}^{a_f}  \frac{ \textcolor {red}{ a^5}  F(a) }{  H(a)}
	\left[f_L(a)+f_H(a)\textcolor {red}{+ f_g(a)} \right]
	\, \mathrm{d} \, a
	\mbox{  }
	\label{Rotate.6.C.I}
\end{eqnarray}
and
\begin{eqnarray}
	\hspace{-2 cm}
D_{II}(a_f) 
= \frac{  H_0^2 \Omega_\Lambda}{6 H_f a_f^3}  \left(\frac{2\pi}{3} \alpha_3 - \frac{ 
\alpha_2}{12}  \right)\textcolor {red}{\frac{1}{a_f^3}} \int_{a_i}^{a_f}  \frac{  \textcolor {red}{ a^5}    }{  H(a)} [f_{HH}(a)\textcolor {red}{+ f_{gg}(a)+ f_g(a)f_H(a)}]
\, \mathrm{d} \, a
&&	\nonumber \\ 
	\hspace{-3 cm}
+ \frac{\pi  H_f}{27 \textcolor {red}{a_f^6}} 
\int_{a_i}^{a_f} \frac{   \textcolor {red}{ a^5} F(a) }{  H(a)}
\left[f_L(a)f_H(a) + f_{LL}(a) +f_{HH}(a)\textcolor {red}{+ f_{gg}(a)+ f_g(a)f_H(a)+ f_g(a)f_L(a)}\right]
\, \mathrm{d} \, a
&&	\nonumber \\ 
\mbox{  }&&
\label{Rotate.6.C.II}
\end{eqnarray}
are dimensionless parameters, and the various functions and parameters in (\ref{Rotate.6.C.I}) and (\ref{Rotate.6.C.II}) are defined in the appendices.  

Evaluating the integrals in (\ref{Rotate.6.C.I}) gives the formulas for $D_I(a)$ for all cosmological eras from the end of inflation until the present time.  The formulas are given in \ref{Rotate.correct} including inflation or in \ref{Rotate.noinflation} neglecting inflation.  The formulas include the effect that with rotation, flow lines are not normal to surfaces of constant global time.  

\section{\label{Rotate.Saddlepoint}      Saddlepoint approximation}

Equation 
(\ref{Rotate.small.average.omega}) shows that (\ref{Rotate.relative-acceleration0c}) has a saddlepoint at $\langle {\omega}_f\rangle =0$.  The path integral in (\ref{Rotate.relative-acceleration0c}) can be approximated by a saddlepoint integration to give
\begin{eqnarray}
	&&
	\psi_f = A(0)
	{{\omega}_m \sqrt{\pi}}/{\sqrt{D_I(a_f)}} e^{i\pi/4}
	\nonumber \\ &&
	\mbox{ for }
	\langle {\omega}_f\rangle 
	< 
	\frac{{\omega}_m}
	{\sqrt{\mid D_I(a_f)\mid}} 
	\approx {T^*}\sqrt{\frac{H_fc^3}{r_0^3a_f^{\textcolor {blue}{6}}}}\frac{1}{ \sqrt{\mid D_I(a_f)\mid}} 
	\mbox{, } 
	\nonumber \\ 
	&&
	\psi_f \approx 0\mbox{, otherwise. }
	\label{Rotate.integration4b.2dup.2b}
\end{eqnarray}

Equation 
(\ref{Rotate.integration4b.2dup.2b}) suggests that ${\omega}_m$ given by (\ref{Rotate.omegaSub.m.2}) gives the main contribution to the limit on rotation rate.  That is verified by the calculation of $D_I(a_f)$ in \ref{Rotate.correct} including inflation or in \ref{Rotate.noinflation} neglecting inflation.  Calculating the formulas for $D_I(a_f)$ was straightforward, although tedious.  The details for that derivation are left out for brevity\textcolor {red}{, but the equations in the appendices are sufficient to allow one to make that derivation.}  
 
\section[      Maximum rotation]{\label{Rotate.results}      Maximum large-scale rotation as a function of cosmic time}

Tables \ref{Rotate.T6} and \ref{Rotate.T5} give the results.  Relative rotation varies as $a^{-1}$ in the radiation era and as $a^{-2}$ in the matter era, where $a$ is the cosmological scale factor \citep[Table 6.1]{Ellis-Maartens-MacCallum:2012}.  Therefore, the relevant results for relative rotation in tables \ref{Rotate.T6} and \ref{Rotate.T5} are for $m_r=1$ and $m_m=2$.  Other values of $m_r$ and $m_m$ are included for comparison.  It is estimated that there were about 50 to 60 e-foldings during inflation \cite{Planck:Collaboration:2018:VI:publ}.  Therefore, table \ref{Rotate.T6} includes calculations for 50, 55, and 60 e-foldings.  

\textcolor {red}{As can be seen from table \ref{Rotate.T6}, even as early as about $10^{-11}$ seconds after the initial singularity, relative rotation rate was limited to be less than about} \textcolor {blue}{$10^{-31}$ rad yr$^{-1}$} \textcolor {red}{for 50 e-foldings and $m_r=1$, and even smaller for 55 and 60 e-foldings or larger values for $m_r$.  A fraction of a second later (about a quarter of a second after the initial singularity), the relative rotation rate is restricted to be less than about} \textcolor {blue}{$2\times 10^{-51}$ rad yr$^{-1}$} \textcolor {red}{for 50 e-foldings.  Notice that the large values for the maximum rotation rate for $m_r=1$ at the end of inflation indicate that the calculation for those values is not valid.}  

\textcolor {red}{However, even without inflation, table \ref{Rotate.T5} shows that relative rotation rate would have been limited to be less than about $5\times 10^{-33}$ rad yr$^{-1}$ at about a quarter of a second after the initial singularity for $m_r=1$.}  The relative rotation rate is limited to even smaller values for later cosmic times.

When calculating the action, in addition to integrating the Lagrangian over a 4-volume, there is also a surface term.  Including the surface term, however, made no significant contribution to the results for the cases in tables \ref{Rotate.T6} and \ref{Rotate.T5}.

\begin{table}[h]
\begin{minipage}{13.5 cm}
\caption{\label{Rotate.T6}Maximum rms rotation rate $\langle\omega_f\rangle_{\mbox{max}}$ as a function of global time $t_f$, $m_r$, and $m_m$ for an inflation era with 50, 55, and 60 e-foldings.}
\begin{tabular}{@{}llllll@{}}
\br
\rule[-1 mm]{0 mm}{4 mm}
$t_f$\footnote{\textcolor {red}{The first column is calculated using equation (\ref{Rotate.t1})}} & $m_r$\footnote{Rotation rate varies as $a^{-m_r}$ in the radiation era.} & $m_m$\footnote{Rotation rate varies as  $a^{-m_m}$ in the matter era.} &$\langle\omega_f\rangle_{\mbox{max}}$ \footnote{The last 3 columns are calculated using equations (\ref{Rotate.integration4b.2dup.2b}), (\ref{Rotate.6.C.I.a.N}), and  (\ref{Rotate.6.C.I.a.6}).} 
&${\langle\omega_f\rangle_{\mbox{max}}}$ 
&${\langle\omega_f\rangle_{\mbox{max}}}$    \\
 &&&for 50 e-foldings&for 55 e-foldings&for 60 e-foldings\\
&&&rad yr$^{-1}$&rad yr$^{-1}$&rad yr$^{-1}$\\
\mr
start of inflation&&&&&\\
 $ 10^{-36}$ s&&&&&\\
\mr
end &1&& $  1 \times 10^{17}$& $  7 \times 10^{14}$& $  7 \times 10^{12}$ \\
of &2&& $5 \times 10^{-7}$& $2 \times 10^{-11}$& $1 \times 10^{-13}$\\
inflation &3&& $ 2 \times 10^{-37}$& $ 8 \times 10^{-46}$& $ 2 \times 10^{-53}$  \\
$ 10^{-34}$ s &4 &&$3 \times 10^{-60}$ &$1 \times 10^{-69}$  &$2 \times 10^{-79}$  \\
\mr
&1&& $ 1 \times 10^{-31}$ & $ 1 \times 10^{-33}$ & $ 8 \times 10^{-36}$    \\
 &2&&$ 7 \times 10^{- 55 }$&$ 3 \times 10^{-59}$&$ 2 \times 10^{-63}$   \\
 &3&&$ 3 \times 10^{-86}$&$ 9 \times 10^{-94}$&$ 2 \times 10^{-101}$ \\
$2.4\times 10^{-11}$ s&4&&$ 6 \times 10^{- 108}$&$ 2 \times 10^{-117}$&$ 3 \times 10^{-127}$\\
\mr
&1&& $ 2 \times 10^{-51}$& $ 5 \times 10^{- 57 }$& $ 2 \times 10^{- 60 }$  \\
radiation &2&&$ 7  \times 10^{-79}$&$ 5  \times 10^{- 83 }$&$ 1  \times 10^{- 87 }$  \\
era &3&&$ 1  \times 10^{- 110 }$&$ 2  \times 10^{- 118 }$ &$ 5  \times 10^{- 126 }$ \\
0.24 s&4&&$ 5  \times 10^{- 142}$&$ 2  \times 10^{- 150}$&$ 3  \times 10^{- 159}$ \\
\mr
matter/ &1&& $ 3  \times 10^{-76 }$& $ 7  \times 10^{-82 }$& $ 5  \times 10^{-97 }$  \\
radiation &2&&$ 7  \times 10^{- 104 }$&$ 7  \times 10^{- 108 }$&$ 2  \times 10^{- 124 }$ \\
equality &3&&$ 8  \times 10^{- 142 }$&$ 2  \times 10^{- 148 }$ &$ 8  \times 10^{- 155 }$ \\
$5\times 10^4$ yr &4&&$8 \times 10^{- 180 }$&$1 \times 10^{- 188 }$&$2 \times 10^{- 197 }$ \\
\mr
&1&2&$ 4  \times 10^{- 77 }$&$ 3  \times 10^{- 81 }$ &$ 2  \times 10^{- 81 }$  \\
recombi-&2&2&$ 2  \times 10^{- 100 }$&$ 8  \times 10^{- 105 }$&$ 4  \times 10^{- 139 }$  \\
nation &3&3&$ \approx 10^{- 138 }$&$ \approx 10^{- 144 }$ &$ \approx 10^{- 138 }$ \\
$3.8\times 10^{5}$ yr &4&4&$ \approx 10^{- 176 }$&$ \approx 10^{- 185 }$&$ \approx 10^{- 194 }$\\
\mr
matter/ &1&2&$ 5  \times 10^{- 90 }$&$ 4  \times 10^{- 94 }$&$ 2  \times 10^{- 94 }$   \\
dark energy &2&2&$ 2  \times 10^{- 113  }$&$ 8  \times 10^{- 118  }$&$ 6  \times 10^{- 122  }$   \\
equality&3&3&$  \approx 10^{- 156 }$&$  \approx 10^{- 163 }$&$  \approx 10^{- 169 }$\\
$10\times 10^{9}$ yr  &4&4&$  \approx 10^{- 197 }$&$  \approx 10^{- 206 }$ &$  \approx 10^{- 215 }$  \\
\mr
&1&2&$ 2 \times 10^{- 90 }$&$ 1 \times 10^{- 94 }$&$ 6 \times 10^{- 95 }$   \\
today &2&2&$ 7  \times 10^{- 114 }$&$ 5  \times 10^{- 118 }$&$ 2  \times 10^{- 122 }$ \\
&3&3&$  \approx 10^{- 156 }$&$  \approx 10^{- 163 }$&$  \approx 10^{- 169 }$ \\
$13.8\times 10^{9}$ yr &4&4&$  \approx 10^{- 197}$&$  \approx 10^{- 206}$ &$  \approx 10^{- 215}$\\
\br
\end{tabular}
\end{minipage}
\end{table}

\begin{table}[h]
\begin{minipage}{13.5 cm}
\caption{\label{Rotate.T5}Cosmological scale factor $a_f$, Hubble parameter $H_f$, and maximum rms rotation rate $\langle\omega_f\rangle_{\mbox{max}}$ as a function of global time $t_f$, $m_r$, and $m_m$ neglecting inflation. 
$r_0 \approx 46.5 \times 10^9$ light years. 
$T^*\approx 1.7\times 10^{-51}$ yr is the Planck time.}
\begin{tabular}{@{}lllllll@{}}
\br
\rule[-1 mm]{0 mm}{5 mm}
$t_f$\footnote{\textcolor {red}{The first column is calculated using equation (\ref{Rotate.t1})}} &   $a_f$ & $H_f$\footnote{\textcolor {red}{The third column is calculated using equation (\ref{Rotate.H})}}  &$T^*\sqrt{\frac{H_f c^3}{(r_0 a_f^2 )^3}}$& $m_r$\footnote{Rotation rate varies as $a^{-m_r}$ in the radiation era.}
 & $m_m$\footnote{Rotation rate varies as $a^{-m_m}$ in the matter era.} &$\langle\omega_f\rangle_{\mbox{max}}$ \footnote{The last column is calculated using equations (\ref{Rotate.integration4b.2dup.2b}) and  (\ref{Rotate.6.C.I.no.inf.5}).}
 \\
&&s$^{-1}$&rad yr$^{-1}$&&&rad yr$^{-1}$\\
\mr
end of inflation &$a_N=$&&& &&  \\
$ 10^{-34}$ s &   $2\times 10^{-27}$ & $5\times 10^{33}$ &$8.5\times 10^{33}$& && \\
\mr
&&&&1&& $ 3 \times 10^{-13}$  \\
 &&&&2&&$ 1 \times 10^{-14}$  \\
 &&&&3&&$ 2 \times 10^{-18}$ \\
$2.4\times 10^{-11}$ s&$10^{-15}$&$2\times 10^{10}$ &$1.3\times 10^{-13}$&4&&$ 6 \times 10^{-23}$ \\
\mr
&&&&1&& $ 5 \times 10^{- 33 }$  \\
radiation &&&&2&&$ 6  \times 10^{- 35 }$  \\
era &&&&3&&$ 3  \times 10^{- 43 }$ \\
0.24 s&$10^{-10}$&$2$&$1.4\times 10^{-33}$&4&&$ 7  \times 10^{- 53}$ \\
\mr
matter/ &&&&1&& $ 8  \times 10^{-58 }$  \\
radiation &&&&2&&$ 6  \times 10^{- 60 }$  \\
equality &$a_{\mbox{eq}}=$&&&3&&$ 1  \times 10^{- 74 }$ \\
$5\times 10^4$ yr & $3\times 10^{-4}$ & $3\times 10^{-13}$ &$1.9\times 10^{-58}$&4&&$1 \times 10^{- 90 }$ \\
\mr
&&&&1&2&$ 7  \times 10^{- 60 }$ \\
recombi-&&&&2&2&$ 1  \times 10^{- 60 }$ \\
nation &&&&3&3&$ 2 \times 10^{- 76 }$ \\
$3.8\times 10^{5}$ yr & $.9\times 10^{-3}$ & $5\times 10^{-14}$ &$2.9\times 10^{-59}$&4&4&$ 7 \times 10^{- 93 }$ \\
\mr
matter/ &&&&1&2&$ 9  \times 10^{- 75 }$  \\
dark energy &&&&2&2&$ 8  \times 10^{-76  }$  \\
equality&&&&3&3&$  8 \times  10^{- 94 }$ \\
$10\times 10^{9}$ yr  & $a_\Lambda = 0.76$ & $2.5\times 10^{-18}$ &$3.3\times 10^{-72}$&4&4&$  3 \times  10^{- 113 }$ \\
\mr
&&&&1&2&$ 4 \times 10^{- 75 }$  \\
today &&&&2&2&$ 3  \times 10^{- 76 }$  \\
&&&&3&3&$  3 \times  10^{- 94 }$ \\
$13.8\times 10^{9}$ yr & $a_0=1.0$ & $2.2\times 10^{-18}$ &$1.4\times 10^{-72}$&4&4&$ 6 \times  10^{- 114}$ \\
\br
\end{tabular}
\end{minipage}
\end{table}

\section{\label{Rotate.Disc}      Discussion}

Table \ref{Rotate.T6} shows that the only cosmologies that contributed significantly to the path integral had very small relative rotation rates already \textcolor {red}{at a fraction of a second after the initial singularity.}  Table \ref{Rotate.T5} shows that even though inflation dominated the calculation, even if there had been no inflation, the only cosmologies that would have contributed significantly to the path integral would have had very small relative rotation rates already within the first second after the initial singularity.

Notice that the main results are due to the small value of ${\omega}_m$ in (\ref{Rotate.omegaSub.m.2}), as can be seen in (\ref{Rotate.integration4b.2dup.2b}).  The value of ${\omega}_m$, which is given in column 4 in table \ref{Rotate.T5}, depends only on Planck's constant, Newton's gravitational constant, the speed of light, the present size of the visible universe, the cosmological scale factor (column 2 in table \ref{Rotate.T5}), and the Hubble parameter (column 3 in table \ref{Rotate.T5}).  The final result is further enhanced by the Factor $D_I$, as can be seen in (\ref{Rotate.integration4b.2dup.2b}).  The further enhancement is a large amount when inflation is included, as can be seen in table \ref{Rotate.T6}.  The enhancement is less when neglecting inflation, as can be seen in table \ref{Rotate.T5}).

\section*{Data availability statement}

No new data were created or analyzed in this study.

\ack
I thank David Peterson and David Bartlett for useful discussion and for helpful suggestions on improving the manuscript. 

\appendix

\section{\label{Rotate.simple}      Background cosmology}

The Hubble parameter, neglecting vorticity, shear, and acceleration, is
\begin{equation}
	\frac{1}{a}\frac{da}{dt} = H(a)= H_0\sqrt{ \Omega_\Lambda + \frac{\Omega_m}{a^3} + \frac{\Omega_r}{a^4}+ \frac{\Omega_k}{a^2}} 
	=\sqrt{\frac{\Lambda}{3} + \frac{8\pi\rho}{3}-\frac{k}{a^2}}
	\mbox{ , }
	\label{Rotate.H}
\end{equation}
where $\Lambda$ is the cosmological constant, $\rho$ is density, 
$t$ is global time, $H_0 = 67.66$ km s$^{-1}$ Mpc$^{-1} = 6.92 \times 10^{-11}$ yr$^{-1}$ $\approx 2.193 \times 10^{-18}$ s$^{-1}$ \cite{Planck:Collaboration:2018:VI:publ} is the present value of the Hubble parameter, $a=1/(z+1)$ is the cosmological scale factor, whose present value is 1, $z$ is the redshift factor, and $\Omega_\Lambda = 0.6889$ \cite{Planck:Collaboration:2018:VI:publ} is the dark energy density divided by the critical density today.  Using $z_{\mbox{eq}} = 3387$ \cite{Planck:Collaboration:2018:VI:publ} for the redshift at radiation/matter equality with $\Omega_m = 0.311$ \cite{Planck:Collaboration:2018:VI:publ} for the matter density today divided by the critical density gives $\Omega_r = 9.181 \times 10^{-5} \approx 9 \times 10^{-5}$ for the radiation energy density divided by the critical density today.\footnote{The values are from \cite[Table 2]{Planck:Collaboration:2018:VI:publ}, and include lensing and baryon acoustic oscillations.}  $\Omega_k = 0.0$ within measurement error \cite{Planck:Collaboration:2018:VI:publ}.  Equation 
(\ref{Rotate.H}) is not valid during inflation.

For an equation of state, we take 
\begin{equation}
	p=w\rho
	\mbox{ , }
	\label{Rotate.p.w.rho}
\end{equation}
where $p$ is pressure, $\rho$ is density, $w=1/3$ in the radiation-dominated era, and $w=0$ in the matter-dominated era. Whenever $w$ appears in any equation here, it will be for the radiation era, where $w=1/3$.  

The variation of density $\rho$ with cosmological scale factor $a$ is given by \citep[Table 6.1]{Ellis-Maartens-MacCallum:2012} 
\begin{equation}
	\rho = \rho_{eq} (a/a_{eq})^{-3(1+w)} 
	\mbox{ , }
	\label{Rotate.rho.of.r}
\end{equation}
where $\rho_{eq}$ is the value of $\rho$ at the boundary between the radiation era and the matter era where $a=a_{eq}$.  

From (\ref{Rotate.H}), we have
\begin{equation}
	\rho   = \frac{3 H_0^2}{8\pi} \left( \frac{\Omega_m}{a^3} + \frac{\Omega_r}{a^4} \right)
	\mbox{ , }
	\label{Rotate.rho.smooth}
\end{equation}
which gives a smooth transition between the radiation era and the matter era instead of the abrupt transition given by (\ref{Rotate.rho.of.r}).  We can have a smooth transition for pressure, also, by taking $w$ in (\ref{Rotate.p.w.rho}) to be given by
\begin{equation}
	w   = \frac{1}{3} \left( 1 + \frac{\Omega_r}{\Omega_m} a \right)^{-n}
	\mbox{ , }
	\label{Rotate.w.smooth}
\end{equation}
where $n$ is a positive integer.  The larger $n$ is, the sharper will be the transition.  However, for calculating the action, the result does not depend strongly on how smooth or sharp is the transition.  Therefore, to keep the calculations simple, we take $n=1$ to give
\begin{equation}
	w   = \frac{1}{3} \left( 1 + \frac{\Omega_r}{\Omega_m} a \right)^{-1}
	\mbox{ . }
	\label{Rotate.w.smooth.1}
\end{equation}
Putting (\ref{Rotate.w.smooth.1}) and (\ref{Rotate.rho.smooth}) in (\ref{Rotate.p.w.rho}) gives
\begin{equation}
	p   = \frac{1}{3} \frac{3 H_0^2}{8\pi}  \frac{\Omega_r}{a^4}
	\mbox{ . }
	\label{Rotate.p.smooth.1}
\end{equation}
Putting (\ref{Rotate.p.smooth.1}) and (\ref{Rotate.rho.smooth}) into (\ref{Rotate.L.tilde}) gives 
\begin{equation}
	{L} = \alpha_1 p + \alpha_2 \rho + \alpha_3 \Lambda  = \frac{3 H_0^2}{8\pi} \left[ 8\pi\alpha_3\Omega_\Lambda + \alpha_2 \frac{\Omega_m}{a^3} + \left(\alpha_1 w + \alpha_2 \right) \frac{\Omega_r}{a^4} \right]
	\mbox{ , }
	\label{Rotate.L.tilde.smooth}
\end{equation}
with $w=1/3$. We can approximate (\ref{Rotate.L.tilde.smooth}) in different eras.  
\begin{eqnarray}
	&&
	{L} = \frac{3 H_0^2}{8\pi} \left[  \alpha_2 \frac{\Omega_m}{a^3} + \left(\alpha_1 w + \alpha_2 \right) \frac{\Omega_r}{a^4} \right]
	\mbox{ for $a \le a_m \approx 10^{-2}$ }
	\nonumber \\ &&
	{L} = \frac{3 H_0^2}{8\pi} \left[ 8\pi\alpha_3\Omega_\Lambda + \alpha_2 \frac{\Omega_m}{a^3}  \right]
	\mbox{ for $ a \ge a_m \approx 10^{-2}$ }
	\label{Rotate.L.tilde.smooth.eras}
\end{eqnarray}

We can convert (\ref{Rotate.H}) into an integral to get
\begin{equation}
	t = \frac{1}{H_0} \int_0^a \frac{da}{\sqrt{ \Omega_\Lambda a^2 + \Omega_m/a + \Omega_r/a^2}} 
	\mbox{ . }
	\label{Rotate.t}
\end{equation}
Equation 
(\ref{Rotate.t}) is a well-defined integral to give the global time $t$ as a function of the cosmological scale factor $a$.  Although it is not easy to calculate in closed form, there is no region where more than two terms in the radical are significant.  That allows a very good approximate evaluation of the integral in closed form.  We have
\begin{eqnarray}
	&& t = \frac{2}{3H_0} \frac{\Omega_r^{3/2}}{\Omega_m^2} \left[2-\left(2-\frac{\Omega_m}{\Omega_r}a\right) \sqrt{1+ \frac{\Omega_m}{\Omega_r}a}\right] 
	\mbox{ for $a\le a_m \approx 10^{-2}$, }
	\nonumber \\ &&
	\mbox{ and }
	\nonumber \\ &&
	t=\frac{1}{3H_0 \sqrt{\Omega_\Lambda}} \ln{\frac{\sqrt{1+\frac{\Omega_m}{\Omega_\Lambda}a^{-3}}+1}{\sqrt{1+\frac{\Omega_m}{\Omega_\Lambda}a^{-3}}-1}}
	\mbox{ for $a\ge a_m$, }
	\label{Rotate.t1}
\end{eqnarray}
where
\begin{equation}
	a_{\mbox{eq}}=\frac{\Omega_r}{\Omega_m} \approx 3\times 10^{-4}   \ll a_m \ll   \left(\frac{\Omega_m}{\Omega_\Lambda} \right)^{1/3}   \approx 0.76
	\mbox{ . }
	\label{Rotate.a-m}
\end{equation}

Using (\ref{Rotate.H}) and (\ref{Rotate.t1}) gives $H$ as a function of global time $t$ in table \ref{Rotate.T5}.  

When the cosmological scale factor $a$ is very small, we can make some approximations that are valid in the very early universe.
\begin{equation}
	a^2 \approx 2 H_0 \sqrt{\Omega_r} t  
	\mbox{, that is, } t \approx 2.4 \times 10^{19} a^2 
	\mbox{ seconds } 
	\mbox{ for $a\ll \Omega_r/\Omega_m$. }
	\label{Rotate.a.small}
\end{equation}
\begin{equation}
	H \approx \frac{1}{2t} 
	\mbox{ for $a\ll \Omega_r/\Omega_m$, but not in the inflation era.}
	\label{Rotate.H.for.a.small}
\end{equation}

For the inflation era (from $10^{-36}$ seconds to $10^{-34}$ seconds), we choose a constant value for the Hubble parameter $H$ that will give 50, 55, or 60 e-foldings.  It is estimated that there were about 50 to 60 e-foldings during inflation \cite{Planck:Collaboration:2018:VI:publ}.  
\textcolor {red}{This gives
\begin{equation}
H(a) = \frac{N}{t_N-t_i} \approx \frac{N}{t_N} 
\mbox{ for $t_i \le t \le t_N$, }
\label{Rotate.H.Inflation}
\end{equation}
where $N$ is the number e-foldings during inflation.}

\section{\label{Rotate.main}   \,\, Action}

We can write the action as
\begin{equation}
	I = 
	\int_{t_i}^{t_f} \int (-g^{(4)})^{1/2} {L}  \mathrm{d}^3x \mathrm{d}t + \mbox{surface term}
	\mbox{ , }
	\label{Rotate.6.1a}
\end{equation}
where $t_i$ is the initial time (which we take to be the beginning of the inflation era), $t_f$ is the final time, which we choose arbitrarily to calculate the action at any specified cosmic time, and $L$ is the Lagrangian.  The surface term is necessary to insure consistency if the action integral is broken into parts \citep{York72,Hawking79}. 

We can convert the time integral in (\ref{Rotate.6.1a}) to an integral over the cosmological scale factor $a$
\begin{equation}
	I =  \int_{a_i}^{a_f} \int \frac{(-g^{(4)})^{1/2} {L}  \mathrm{d}^3x \, \mathrm{d} \, a }{\dot{a}}
	+ \overline{\frac{1}{2} r_0^3 a_i^2 \dot{a}_i} -  \overline{\frac{1}{2} r_0^3 a_f^2 \dot{a}_f}
	\mbox{ , }
	\label{Rotate.6.1b}
\end{equation}
where $\dot{a}=\mathrm{d}a/\mathrm{d}t$ is given by a generalization of the Friedmann equation that includes rotation  \cite{Jones:rotation:problem:2020}, $a_i$ is the value of the cosmological scale factor at $t=t_i$, $a_f$ is the value of the cosmological scale factor at $t=t_f$, the surface terms at the initial and final times have been included, $r_0$ is the present radius of the cosmological horizon, and an overbar indicates a spatial average.

\color {red} We can express the volume integral in (\ref{Rotate.6.1b}) as a product of the spatial volume and the spatial average.  
\begin{equation}
	I 
	= \int_{a_i}^{a_f}  V(a)
	\overline{(-g^{(4)})^{1/2}\left( \frac{  {L} }{\dot{a}} \right) }
	\, \mathrm{d} \, a 
	+ \overline{\frac{1}{2} r_0^3 a_i^2 \dot{a}_i} -  \overline{\frac{1}{2} r_0^3 a_f^2 \dot{a}_f}
	\mbox{ , }
	\label{Rotate.6.1.c.1a}
\end{equation}
where we have used the knowledge that our universe is spatially flat, 
\begin{equation} V(a) = \frac{4}{3} \pi a^3 r_0^3 \mbox{  } \label{Rotate.volume} \end{equation} 
is the approximate spatial volume,\footnote{\textcolor {red}{Since our universe is approximately spatially flat.  Although there are arguments that $V$ should be infinite because this is an open cosmology \citep{Hartle2005}, considering causality leads to restricting the spatial part of the action to }\textcolor {blue}{a finite value.}} 
and $r_0$ is the present radius of the cosmological horizon.  

The FLRW metric for the spatially flat case generalized to include vorticity and shear is
\begin{eqnarray} 
	&
	{g_{\mu\nu}} = 
	\left( \begin{array}{cccc}
		-1 &
	a^2	(w_1+\tilde{w_1}) &
		a^2	(w_2+\tilde{w_2}) &
		a^2	(w_3+\tilde{w_3})\\
		a^2(w_1+\tilde{w_1}) &
		a^2 &
		0 &
		0 \\
		a^2	(w_2+\tilde{w_2}) &
		0 &
		a^2 &
		0 \\
		a^2	(w_3+\tilde{w_3}) &
		0 &
		0 &
		a^2	
	\end{array} \right) 
	, 
	&
	\label{Rotate.gmunu}
\end{eqnarray}
where (using the notation of \citep[Chapter 29]{Hamilton:2023}) the $w_i$ and the $\tilde{w_i}$ are parameters representing the vector mode, that includes vorticity and shear.  Since vorticity and shear are coupled, it is necessary to include both.  The components of the vorticity are $\omega_1=\overline{\omega}_{23}$, $\omega_2=\overline{\omega}_{31}$, $\omega_3=\overline{\omega}_{12}$, where
\begin{equation}
\overline{\omega}_{ij}=\frac{1}{2a}(\nabla_i \tilde{w}_i-\nabla_j \tilde{w}_i) =-\frac{i}{2a}(k_i \tilde{w_j}-k_j \tilde{w}_i)
\mbox{ . }
\label{Rotate.vorticity}
\end{equation}
The shear is given by
\begin{equation}
	{\sigma}_{ij}=\frac{1}{2a}(\nabla_i {w_j}+\nabla_j {w_i}) =-\frac{i}{2a}(k_i {w_j}+k_j {w_i})
	\mbox{ . }
	\label{Rotate.shear}
\end{equation}
From (\ref{Rotate.gmunu}), we have
\begin{eqnarray}
	&
(-g^{(4)})^{1/2} = a^3 \{1+a^2[(w_1+\tilde{w_1})^2+(w_2+\tilde{w_2})^2+(w_3+\tilde{w_3})^2]\}^{1/2}
	& \nonumber \\ & 
	\approx a^3 (1+a^2\tilde{w}^2)^{1/2}
	\approx a^3 (1-\frac{a^4\omega^2}{k^2})^{1/2}
	\approx a^3 (1-\frac{a^4\omega^2}{2k^2}-\frac{a^8\omega^4}{8k^4})
	& \nonumber \\ & 
	\approx a^3[1+f_g(a)\epsilon+f_{gg}(a)\epsilon^2]
	\mbox{ , }
	&
\label{Rotate.sqrt.g}
\end{eqnarray}
where
\begin{equation}
f_g(a)\equiv -3 \frac{a^4H_f^2}{k^2}F_{\omega}(a)^2
	\mbox{ , }
\label{Rotate.f.g}
\end{equation}
\begin{equation}
	f_{gg}(a)\equiv -6 \frac{a^4H_f^2}{k^2}F_{\omega}(a)^2 g_{\omega}(a)  - \frac{9a^8H_f^4}{2k^4}F_{\omega}(a)^4
	\mbox{ , }
	\label{Rotate.f.gg}
\end{equation}
we neglect the coupling between vorticity and shear, we consider only the magnitude of the vorticity, and we expand for small vorticity.  Neglecting the coupling between vorticity and shear might lead to an error of a few factors of 10 or so, which is allowable in this order-of-magnitude calculation.
\normalcolor

\section{\label{Rotate.Lagrangian.App}   \,\, Approximate {Lagrangian}}

The total Lagrangian is 
\begin{equation}
L= L_{\mbox{geom}} + L_{\mbox{matter}}
\mbox{ , }
\label{Rotate.Lagrangian}
\end{equation}
where
\begin{equation}
L_{\mbox{geom}} = \frac{R-2\Lambda}{16\pi}
\mbox{ , }
\label{Rotate.L.geom}
\end{equation}
is the Lagrangian for the geometry, $R$ is the Ricci scalar, $\Lambda$ is the cosmological constant, and $L_{\mbox{matter}}$ is the matter Lagrangian.  For solutions to Einstein's field equations, the Ricci scalar is given by 
\begin{equation}
R=-8\pi T + 4 \Lambda
\mbox{ , }
\label{Rotate.Ricci}
\end{equation}
where $T$ is the contracted stress-energy tensor.  For a perfect fluid, we have 
\begin{equation}
T=3p-\rho
\mbox{ , }
\label{Rotate.T}
\end{equation}
where $p$ is pressure and $\rho$ is density.  The matter tensor for a perfect fluid is \citep{MacCallum-Taub:1972,Schutz76} 
\begin{equation}
L_{\mbox{matter}} = p
\mbox{ , }
\label{Rotate.L.matter}
\end{equation}
Combining the above, we have
\begin{equation}
L = -\frac{1}{2}p + \frac{1}{2}\rho + \frac{\Lambda}{8\pi}
\mbox{ , }
\label{Rotate.L.total}
\end{equation}
for a perfect fluid for solutions to Einstein's field equations. 

However, to consider more general possibilities, we take
\begin{equation}
{L} = \alpha_1 p + \alpha_2 \rho + \alpha_3 \Lambda
\mbox{ , }
\label{Rotate.L.tilde}
\end{equation}
where $\alpha_1$, $\alpha_2$, and $\alpha_3$ are dimensionless constants of order unity. 

When there is rotation, the scale factor $\ell$ along lines of cosmic flow differs from the cosmological scale factor $a$.  In that case, the effective Lagrangian is 
\begin{equation}
	{L} = \alpha_1 p + \alpha_2 \rho + \alpha_3 \Lambda  \approx \frac{3 H_0^2}{8\pi} \left[ 8\pi\alpha_3\Omega_\Lambda + \alpha_2 \frac{\Omega_m}{\ell^3} + \left(\alpha_1 w + \alpha_2 \right) \frac{\Omega_r}{\ell^4} \right]
	\mbox{ , }
	\label{Rotate.L.tilde.smooth.ell}
\end{equation}
where $\ell$ is not normal to surfaces of constant global time.  However, since we are considering the case where the relative rotation is small, we can expand (\ref{Rotate.L.tilde.smooth.ell}) to second order in a parameter $\epsilon$ that is proportional to the mean square rotation rate in a specified time slice.  That gives
\begin{equation}
	{L} 
	\approx 3H_0^2\Omega_\Lambda \left(\alpha_3 -\frac{\alpha_2}{8\pi}\right) + 
	H_f^2 F (a) \left[1+ f_L(a) \epsilon + f_{LL}(a) \epsilon^2 \right]
	\mbox{ , }
	\label{Rotate.L.sec.order}
\end{equation}
where $H_f = H(a_f)$, $\epsilon$ is defined in (\ref{Rotate.epsilon}), 

\begin{eqnarray}
	F(a) = \frac{3}{8\pi} (\alpha_1 w + \alpha_2) \frac{H_0^2 \Omega_r}{H_f^2 a^4}
	\, \mbox{for } a \le a_{eq} 
	\nonumber \\
	F(a) = \frac{3}{8\pi} 
	\alpha_2 
	\frac{H(a)^2}{H_f^2}
	\, \mbox{for } a \ge a_{eq} 
	\mbox{ , }
	\label{Rotate.F.of.a}
\end{eqnarray}
$a_{eq}$ is the cosmological scale factor at radiation-matter equality, and
\begin{eqnarray}
	f_L(a) = 
	-4 f_\ell(a) 
	\, \mbox{for } a \le a_{eq} 
	\nonumber \\
	f_L(a) = 
	-\frac{H_0^2}{H(a)^2}  \left[3\frac{\Omega_m}{a^3} + 4 \frac{\Omega_r}{a^4}  \right] 
	f_\ell(a) 
	\, \mbox{for } a \ge a_{eq} 
	\mbox{ , }
	\label{Rotate.F.sub.L.of.a}
\end{eqnarray}

\begin{eqnarray}
	&&
	f_{LL}(a) = \frac{a^4}{\Omega_r}  
	\left[ \left(6\frac{\Omega_m}{a^3} + 10 \frac{\Omega_r}{a^4}\right) f_\ell(a)^2 -\left(3\frac{\Omega_m}{a^3} +4 \frac{\Omega_r}{a^4}\right) f_{\ell\ell}(a) \right] 
	\, \mbox{for } a \le a_{eq} 
	\nonumber \\ &&
	f_{LL}(a) = \frac{H_0^2}{H(a)^2}  
	\left[ \left(6\frac{\Omega_m}{a^3} + 10 \frac{\Omega_r}{a^4}\right) f_\ell(a)^2 -\left(3\frac{\Omega_m}{a^3} +4 \frac{\Omega_r}{a^4}\right) f_{\ell\ell}(a) \right] 
	\nonumber \\ &&
	\, \mbox{for } a \ge a_{eq} 
	\mbox{ , }
	\label{Rotate.F.sub.LL.of.a}
\end{eqnarray}
and $f_\ell(a)$ and $f_{\ell\ell}(a)$ are defined in (\ref{Rotate.f.ell}) and (\ref{Rotate.f.ell.ell}).

\section{\label{Rotate.acceleration}   \,\,    Relative rotation angle}

The time-integral of the rotation rate gives an angle 
\begin{equation}
	\theta = \int_0^t \omega \mathrm{d}t
	= \int_{a_i}^a \frac{\omega}{\dot{a}} \mathrm{d}a
	\approx \int_{a_i}^a \frac{\omega}{aH(a)}\left[1+  f_H(a) \epsilon + f_{HH}(a) \epsilon^2 \right] \mathrm{d}a
	\mbox{ , }
	\label{Rotate.theta}
\end{equation}
where we have used (\ref{Rotate.adot}) for $1/\dot{a}$, and where 
\begin{equation}
	\epsilon \equiv \frac{1}{6} \left(\frac{\omega_f}{H_f}\right)^2 
	\mbox{  }
	\label{Rotate.epsilon}
\end{equation}
is a dimensionless small number, $\omega_f$ is the relative rotation rate (a function of position) on the final time slice (which we choose arbitrarily) and $H_f=H(a_f)$ is the value of the Hubble parameter on the final time slice.  

Using (\ref{Rotate.acceleration.1}) with $m=m_r$ in the radiation era and $m=m_m$ in the matter era gives 
\begin{equation}
	\fl
	\omega \approx \omega_f F_{\omega}(a) \left[1+g_{\omega}(a) \epsilon + g_{\omega\omega}(a) \epsilon^2 \right] 
	\mbox{ , }
	\label{Rotate.acceleration.2}
\end{equation}
\begin{eqnarray}
	\fl
	F_{\omega}(a) = 
	\left(\frac{a_f}{a}\right)^{m_r} 
	\, \mbox{for } a \le a_{eq} \mbox{ and }   a_f \le a_{eq}
	\nonumber \\
	\fl
	F_{\omega}(a) = 
	\left(\frac{a_f}{a_{eq}}\right)^{m_m} 
	\left(\frac{a_{eq}}{a}\right)^{m_r} 
	\, \mbox{for } a \le a_{eq} \mbox { and }   a_f \ge a_{eq}
	\nonumber \\
	\fl
	F_{\omega}(a) = 
	\left(\frac{a_f}{a}\right)^{m_m} 
	\, \mbox{for } a \ge a_{eq} \mbox{ and }   a_f \ge a_{eq}
	\mbox{ , }
	\label{Rotate.F.sub.a}
\end{eqnarray}
\begin{eqnarray}
	&&
	\fl
	g_{\omega}(a) = 
	m_r \left[f_\ell(a_f) - f_\ell(a) \right]
	\, \mbox{for } a \le a_{eq} \mbox{ and }   a_f \le a_{eq}
	\nonumber \\ &&
	\fl
	g_{\omega}(a) = 
	m_mf_\ell(a_f)+ (m_r-m_m) f_\ell(a_{eq})-m_rf_\ell(a)
	\, \mbox{for }  a \le a_{eq} \mbox{ and }   a_f \ge a_{eq}
	\nonumber \\ &&
	\fl
	g_{\omega}(a) = 
	m_m \left[f_\ell(a_f) - f_\ell(a) \right]
	\, \mbox{for }  a \ge a_{eq} \mbox{ and }   a_f \ge a_{eq}
	\mbox{ , }
	\label{Rotate.g.a}
\end{eqnarray}
and
\begin{eqnarray}
	\fl
	g_{\omega\omega}(a) = 
	m_r \left[f_{\ell\ell}(a_f) - f_{\ell\ell}(a) +\frac{m_r-1}{2} f_\ell(a_f)^2 +\frac{m_r+1}{2} f_\ell(a)^2 -m_r f_\ell(a_f)f_\ell(a)  \right]
	\nonumber \\ 
	\fl
	\, \mbox{for } a \le a_{eq} \mbox{ and }   a_f \le a_{eq}
	\nonumber \\ 
	\fl
	g_{\omega\omega}(a) = 
	m_m f_{\ell\ell}(a_f) 
	+ (m_r-m_m) f_{\ell\ell}(a_{eq})
	-m_r f_{\ell\ell}(a)
	\nonumber \\ 
	\fl
	+\frac{m_m(m_m-1)}{2} f_{\ell}(a_f)^2 
	+\frac{(m_r-m_m)(m_r-m_m-1)}{2} f_{\ell}(a_{eq})^2 
	+\frac{m_r(m_r+1)}{2} f_{\ell}(a)^2 
	\nonumber \\ 
	\fl
	+m_m (m_r-m_m) f_{\ell}(a_f)f_{\ell}(a_{eq}) 
	-m_m m_r f_{\ell}(a_f)f_{\ell}(a) 
	-m_r (m_r-m_m) f_{\ell}(a_{eq}) f_{\ell}(a)
	\nonumber \\ 
	\fl
	\, \mbox{for }  a \le a_{eq} \mbox{ and }   a_f \ge a_{eq}
	\nonumber \\ 
	\fl
	g_{\omega\omega}(a) = 
	m_m \left[f_{\ell\ell}(a_f) - f_{\ell\ell}(a) +\frac{m_m-1}{2} f_\ell(a_f)^2 +\frac{m_m+1}{2} f_\ell(a)^2 -m_m f_\ell(a_f)f_\ell(a)  \right]
	\nonumber \\ 
	\fl
	\, \mbox{for }  a \ge a_{eq} \mbox{ and }   a_f \ge a_{eq}
	\mbox{ . }
	\label{Rotate.g.aa}
\end{eqnarray}

Using (\ref{Rotate.acceleration.2}) into (\ref{Rotate.theta}) gives
\begin{equation}
	\theta = \sqrt{2\epsilon} \left[\sqrt{f_\theta(a)} -\frac{1}{2} f_{\theta\theta}(a)\epsilon \right]
	\mbox{ , }
	\label{Rotate.theta.2}
\end{equation}
where
\begin{equation}
	\sqrt{f_\theta(a)} \equiv \sqrt{3} \int_{a_i}^a \frac{H_f}{H(a)} \frac{F_{\omega}(a)}{a} \mathrm{d}a
	\mbox{ , }
	\label{Rotate.f.theta}
\end{equation}
and
\begin{equation}
	{f_{\theta\theta}(a)} \equiv -2\sqrt{3} \int_{a_i}^a \frac{H_f}{H(a)} \frac{F_{\omega}(a)}{a} \left[g_{\omega}(a) + f_H(a) \right] \mathrm{d}a
	\mbox{ . }
	\label{Rotate.f.theta.theta}
\end{equation}

\section{\label{Rotate.ell}  \,\,  Deviation of the local scale factor from the global cosmological scale factor}

With rotation, the scale factor $\ell$ along lines of cosmic flow differs from the global cosmological scale factor $a$.  Thus 
\begin{equation}
	\ell \equiv \int \frac{ \mathrm{d}a}{\cos{\theta}}
	\approx a[1+f_\ell(a) \epsilon +f_{\ell\ell}(a) \epsilon^2]
	\mbox{ , }
	\label{Rotate.ell.1}
\end{equation}
where
\begin{equation}
	f_\ell(a) \equiv \frac{1}{a} \int_{a_i}^a f_\theta(a) \mathrm{d}a
	\mbox{  }
	\label{Rotate.f.ell}
\end{equation}
and
\begin{equation}
	f_{\ell\ell}(a) \equiv \frac{1}{a} \int_{a_i}^a \left[\frac{5}{6} f^2_\theta(a) - \sqrt{f_\theta(a)}f_{\theta\theta}(a) \right] \mathrm{d}a
	\mbox{ . }
	\label{Rotate.f.ell.ell}
\end{equation}

\section{\label{Rotate.Friedmann}   \,\, Approximate Generalized Friedmann equation}

If there were no rotation, then we could use the Friedmann equation to calculate $\dot{a}$ in (\ref{Rotate.6.1b}).  However, with rotation, the Friedmann equation, generalized to include vorticity and shear, can be calculated from the Raychoudhury equation to give \cite{Jones:rotation:problem:2020} 
\begin{equation}
	{\dot{\ell}}
	=  \ell \sqrt{ H(\ell)^2   + H_{\omega}^2 + H_{\sigma}^2 
	}
	\mbox{ , }
	\label{Rotate.Friedmann.1.ell.dot}
\end{equation}
where $\ell$ is a scale factor along lines of cosmic flow.     In the presence of rotation, $a$ and $\ell$ differ.  $H(\ell)$ [given by (\ref{Rotate.H}) in \ref{Rotate.simple} with $a\to \ell$] is the Hubble parameter without vorticity, shear, or acceleration, 
\begin{equation} H_{\omega}^2 \equiv \frac{4}{3\ell^2}    \int \ell  \omega^2  \, \mathrm{d} \ell \mbox{  } \label{Rotate.H.omega.ell} \end{equation}
is the vorticity term, $\omega$ is vorticity, 
\begin{equation} H_{\sigma}^2 \equiv -\frac{4}{3\ell^2}    \int \ell  \sigma^2  \, \mathrm{d} \ell \mbox{  } \label{Rotate.H.sigma.ell} \end{equation}
is the shear term, and $\sigma$ is shear.

We can take rotation rate to depend on the distance along flow lines $\ell$ as 
\begin{equation}
	\omega \propto {\ell}^{-m}
	\mbox{ , }
	\label{Rotate.acceleration.1}
\end{equation}
where the value of $m$ depends on the assumptions we make.  We use $m=m_r$ in the radiation era and $m=m_m$ in the matter era.  We can take $m_r=1$ and $m_m=2$ \citep[Table 6.1]{Ellis-Maartens-MacCallum:2012}.\footnote{Alternatively, we could include coupling between vorticity and shear in terms of a vector perturbation \citep[Chapter 10]{Ellis-Maartens-MacCallum:2012}, but the effect on the final result would not likely be significant for an order-of-magnitude calculation such as this.}  

Using (\ref{Rotate.ell.1}) and (\ref{Rotate.acceleration.1}) in (\ref{Rotate.H.omega.ell}) gives 
\begin{equation}
	H_{\omega}^2 \approx H_f^2\left[ f_{\omega}(a) \epsilon + f_{\omega\omega}(a) \epsilon^2 \right] 
	\mbox{ , }
	\label{Rotate.H.a}
\end{equation}
where
\begin{eqnarray}
	&&
	f_{\omega}(a) = 
	4\left(\frac{a_f}{a}\right)^2 \left[2\ln{\frac{a_f}{a}} +\alpha_4 -1 \right]
	\, \mbox{for } a \le a_{eq} \mbox{ and }   a_f \le a_{eq} \mbox{ and }   m_r=1
	\nonumber \\
	&&
	f_{\omega}(a) = 
	4\left(\frac{a_f}{a}\right)^2 \left[ \alpha_4 -\frac{1}{m_r -1} \left({\frac{a_f}{a}}\right)^{2(m_r-1)}  \right]
	\nonumber \\ &&
	\, \mbox{for }  a \le a_{eq} \mbox{ and }   a_f \le a_{eq}  \mbox{ and }   m_r\ne 1
	\nonumber \\ &&
	f_{\omega}(a) = 
	4\left(\frac{a_f}{a_{eq}}\right)^{2(m_m-m_r)} \left(\frac{a_f}{a}\right)^2 \left[2\ln{\frac{a_f}{a}} +\alpha_4 -1 \right]
	\nonumber \\ &&
	\, \mbox{for }  a \le a_{eq} \mbox{ and }   a_f \ge a_{eq}  \mbox{ and }   m_r=1
	\nonumber \\
	&&
	f_{\omega}(a) = 
	4\left(\frac{a_f}{a_{eq}}\right)^{2(m_m-m_r)} \left(\frac{a_f}{a}\right)^2 \left[ \alpha_4 -\frac{1}{m_r -1} \left({\frac{a_f}{a}}\right)^{2(m_r-1)}   \right]
	\nonumber \\ &&
	\, \mbox{for }  a \le a_{eq} \mbox{ and }   a_f \ge a_{eq}  \mbox{ and }   m_r\ne 1
	\nonumber \\
	&&
	f_{\omega}(a) = 
	4\left(\frac{a_f}{a}\right)^2 \left[ \alpha_4 -\frac{1}{m_m -1} \left({\frac{a_f}{a}}\right)^{2(m_m-1)}   \right]
	\nonumber \\ &&
	\, \mbox{for }  a \ge a_{eq} \mbox{ and }   a_f \ge a_{eq}  \mbox{ and }   m_m\ne 1
	\mbox{ , }
	\label{Rotate.f.a}
\end{eqnarray}
and
\begin{eqnarray}
	&&
	f_{\omega\omega}(a) = 
	8\left(\frac{a_f}{a}\right)^2 
	\left[2\ln{\frac{a_f}{a}} +\alpha_4  \right]
	\left[ f_\ell(a_f) - f_\ell(a) \right]
	\nonumber \\
	&&
	\, \mbox{for } a \le a_{eq} \mbox{ and }   a_f \le a_{eq} \mbox{ and }   m_r=1
	\nonumber \\
	&&
	f_{\omega\omega}(a) = 
	8\left(\frac{a_f}{a}\right)^2 
	\left[ \alpha_4 -\frac{m_r}{m_r -1} \left({\frac{a_f}{a}}\right)^{2(m_r-1)}  \right]
	\left[ f_\ell(a_f) - f_\ell(a) \right]
	\nonumber \\
	&&
	\, \mbox{for }  a \le a_{eq} \mbox{ and }   a_f \le a_{eq}  \mbox{ and }   m_r\ne 1
	\nonumber \\
	&&
	f_{\omega\omega}(a) = 
	8\left(\frac{a_f}{a_{eq}}\right)^{2(m_m-m_r)} 
	\left(\frac{a_f}{a}\right)^2 \left[2\ln{\frac{a_f}{a}} +\alpha_4 + m_m -m_r \right]
	\left[ f_\ell(a_f) - f_\ell(a) \right]
	\nonumber \\
	&&
	\, \mbox{for }  a \le a_{eq} \mbox{ and }   a_f \ge a_{eq}  \mbox{ and }   m_r=1
	\nonumber \\
	&&
	f_{\omega\omega}(a) = 
	8\left(\frac{a_f}{a_{eq}}\right)^{2(m_m-m_r)} \left(\frac{a_f}{a}\right)^2 
	\times
	\nonumber \\ &&
	\left[ \alpha_4 -\frac{m_r}{m_r -1} \left({\frac{a_f}{a}}\right)^{2(m_r-1)}  + m_m -m_r  \right]
	\left[ f_\ell(a_f) - f_\ell(a) \right]
	\nonumber \\
	&&
	\, \mbox{for }  a \le a_{eq} \mbox{ and }   a_f \ge a_{eq}  \mbox{ and }   m_r\ne 1
	\nonumber \\
	&&
	f_{\omega\omega}(a) = 
	8\left(\frac{a_f}{a}\right)^2 
	\left[ \alpha_4 -\frac{m_m}{m_m -1} \left({\frac{a_f}{a}}\right)^{2(m_m-1)}   \right]
	\left[ f_\ell(a_f) - f_\ell(a) \right]
	\nonumber \\
	&&
	\, \mbox{for }  a \ge a_{eq} \mbox{ and }   a_f \ge a_{eq}  \mbox{ and }   m_r\ne 1
	\mbox{ . }
	\label{Rotate.f.aa}
\end{eqnarray}

Using (\ref{Rotate.H.a}) and (\ref{Rotate.ell.1}) in (\ref{Rotate.Friedmann.1.ell.dot}) gives
\begin{equation}
	\frac{1}{\dot{a}} \approx \frac{1}{aH(a)}\left[1+  f_H(a) \epsilon + f_{HH}(a) \epsilon^2 \right] 
	\mbox{ , }
	\label{Rotate.adot}
\end{equation}
where
\begin{equation}
	f_H(a) = -\frac{1}{2}f_L(a)- f_\ell (a)+f_\theta(a)-\frac{1}{2}\frac{H_f^2}{H(a)^2}f_{\omega}(a)
	\mbox{  }
	\label{Rotate.Friedmann.f.H}
\end{equation}
and
\begin{eqnarray}
	&&
	f_{HH}(a) = 
	-\frac{1}{2}f_{LL}(a)
	+ f_{\ell}^2 (a)
	- f_{\ell\ell} (a)
	+\frac{5}{6}f_\theta^2(a)
	-\sqrt{f_\theta(a)}f_{\theta\theta} (a)
	+\frac{3}{8}f_L(a)^2
	\nonumber \\ &&
	+\frac{1}{2}f_L(a)f_\ell (a)
	-\frac{1}{2}f_L(a)f_\theta(a)
	-f_\ell (a)f_\theta(a)
	\nonumber \\ &&  
	+\left(\frac{3}{4}f_L(a)f_{\omega}(a)+\frac{1}{2}f_\ell (a)f_{\omega}(a)-\frac{1}{2}f_\theta(a)f_{\omega}(a) -\frac{1}{2}f_{\omega\omega}(a)
	+\frac{3}{8}\frac{H_f^2}{H(a)^2}f_{\omega}(a)^2
	\right)
	\nonumber \\ &&
	\times \frac{H_f^2}{H(a)^2}
	\mbox{ . }
	\label{Rotate.Friedmann.f.HH}
\end{eqnarray}

\section{\label{Rotate.correct}   \,\, Formulas for $D_I$ including inflation}

\textcolor {red}{We can use  (\ref{Rotate.F.of.a})   for $F(a)$ and (\ref{Rotate.Friedmann.f.H})   for $f_H(a)$ in (\ref{Rotate.6.C.I}) and then use 
 (\ref{Rotate.H}) for $H(a)$ except during inflation,
 (\ref{Rotate.H.Inflation}) for $H(a)$ during inflation,
(\ref{Rotate.p.w.rho}) for $p$,
(\ref{Rotate.rho.of.r}) for $\rho$,
(\ref{Rotate.F.sub.L.of.a}) for $f_L(a)$,  (\ref{Rotate.f.ell}) for $f_\ell(a)$,  (\ref{Rotate.f.theta}) for $f_\theta(a)$, 
(\ref{Rotate.F.sub.a}) for $F_{\omega}(a)$, 
and  (\ref{Rotate.f.a}) for $f_{\omega}(a)$ 
to give the formulas for $D_I$.}  

These formulas are used in table \ref{Rotate.T6}.

At the end of inflation, we have 
\begin{equation}
	D_I(a_N) \approx \left[-\frac{\alpha_1 w + \alpha_2}{\textcolor {blue}{4N^2} } 
		\overbrace{
				+ f_4(a_f)
		}^{\mbox{surface term}} \right]
\frac{e^{2Nm_r}}{4N^2m_r^2}
	\mbox{ , }
	\label{Rotate.6.C.I.a.N}
\end{equation}
where $a_N$ is the cosmological scale factor at the end of inflation and $N$ is the number of e-foldings during inflation.

For $a_f \gg a_N$,  some terms dominate over others, so that we have
\begin{equation}
	D_I(a_f) \approx \left[f(a_f) 
		\overbrace{
			+ f_4(a_f) 
		}^{\mbox{surface term}} \right] \frac{e^{2Nm_r}}{4N^2m_r^2}
	\mbox{ , }
	\label{Rotate.6.C.I.a.6}
\end{equation}
where
\begin{eqnarray}
	&&
	f(a_f) = f_1(a_f) 
	\, \mbox{for } a_N \ll   a_f \le a_{eq} 
	\mbox{ , }
	\nonumber \\ &&
	f(a_f) = f(a_{eq}) + f_2(a_f) f_3(a_f) 
	\, \mbox{for } a_{eq} \le   a_f \le a_{\Lambda},  
	\nonumber \\ &&
	f(a_f) = f(a_{\Lambda}) + f_2(a_f) \textcolor {red}{(\pi \alpha_3-\frac{\alpha_2}{4})}
	\left(\frac{a_f}{a_\Lambda}\right)^{2m_m}
\left[ 
\textcolor {red}{	\frac{1}{3}\left(\frac{a_\Lambda}{a_f}\right)^{3} \left(1- \left(\frac{a_\Lambda}{a_f}\right)^{3} \right)} \right]
	\nonumber \\ &&
	\, \mbox{for } a_{\Lambda} \le   a_f \le a_{0}=1 
	\mbox{ , }
	\label{Rotate.6.C.I.a.6.f}
\end{eqnarray}
$a_{eq}$ is the value of the cosmological scale factor at the time of matter-radiation equality, $a_{\Lambda}$ is the value of the cosmological scale factor at the time of matter-dark energy equality, $a_{0}=1$ is the present value of the cosmological scale factor, 
\begin{eqnarray}
	&&
	f_1(a) \equiv -\frac{\alpha_1 w + \alpha_2}{\textcolor {red}{8} }  \left[\textcolor {blue}{\frac{2}{N^2}} + \left(\frac{a}{a_N}\right)^{2m_r-4} \right] 
	\, \mbox{for } a_{N} \ll   a < a_{eq} 	\mbox{ , }
\nonumber \\ &&
	f_1(a) \equiv -\frac{\alpha_1 w + \alpha_2}{\textcolor {red}{8} }  \left[\textcolor {blue}{\frac{2}{N^2}} + \textcolor {red}{2}\left(\frac{a}{a_N}\right)^{2m_r-4} \right] 
\, \mbox{for }    a = a_{eq} 
	\mbox{ , }
	\label{Rotate.f1}
\end{eqnarray}
\begin{equation}
f_2(a) \equiv 3 \frac{\Omega_m}{\Omega_r}
	\frac{a_N^4}{a^3} 
	\left(\frac{a_{eq}}{a_N}\right)^{2m_r} 
	\left(\frac{a}{a_{eq}}\right)^{2m_m}
	\mbox{ , }
	\label{Rotate.f2}
\end{equation}
\begin{eqnarray}
	&&
	\hspace*{-2.6 cm} f_3(a) \equiv \frac{  \Omega_\Lambda H_0^2}{\textcolor {red}{5} H(a)^2} 
	\left(\frac{2\pi}{3} \alpha_3 - \frac{
		\alpha_2}{12}  \right)  
	\left[1-\left(\frac{a_{eq}}{a}\right)^{\textcolor {red}{15}/2}\right]
	\nonumber \\ &&
	- 	 \frac{\alpha_1 w + \alpha_2}{\textcolor {red}{36} }  
	\left[1- \left(\frac{a_{eq}}{a}\right)^{\textcolor {red}{9}/2} \right]
	\mbox{ , }
	\label{Rotate.f3}
\end{eqnarray}
and
\begin{eqnarray}
	&&
	f_4(a_f) = 
	\mbox{negligible} 
	\, \mbox{for } a_N \ll   a_f \le a_{eq} 
	\mbox{ , }
	\nonumber \\ &&
	f_4(a_f) = - \frac{1}{8\pi}\frac{H_0^2}{H_N^2}\frac{\Omega_m}{a_f^3}
	\left(\frac{a_{eq}}{a_N}\right)^{2m_r}
	\left(\frac{a_f}{a_{eq}}\right)^{2m_m}
	\, \mbox{for }   a_f  \ge   a_{eq}
	\mbox{ . }
	\label{Rotate.6.C.I.a.6.f4}
\end{eqnarray}

The calculations used for table \ref{Rotate.T6} take the quantities $\alpha_1 w$, $\alpha_2$, and $\alpha_3$ to be of order unity.  

\section{\label{Rotate.noinflation} \, \,   Formulas for $D_I$ neglecting inflation}

\textcolor {red}{We can use  (\ref{Rotate.F.of.a})   for $F(a)$ and (\ref{Rotate.Friedmann.f.H})   for $f_H(a)$ in (\ref{Rotate.6.C.I})
and then use 
(\ref{Rotate.H}) for $H(a)$,
(\ref{Rotate.p.w.rho}) for $p$,
(\ref{Rotate.rho.of.r}) for $\rho$,
(\ref{Rotate.F.sub.L.of.a}) for $f_L(a)$,  (\ref{Rotate.f.ell}) for $f_\ell(a)$,  (\ref{Rotate.f.theta}) for $f_\theta(a)$, 
(\ref{Rotate.F.sub.a}) for $F_{\omega}(a)$, 
and  (\ref{Rotate.f.a}) for $f_{\omega}(a)$. 
Keeping only the most significant terms for $a_f \gg a_N$ gives} 
\begin{eqnarray}
	&&
	D_I(a_f) \approx
	\frac{\alpha_1 w + \alpha_2}{\textcolor {blue}{54}} 
		\overbrace{
			-\frac{1}{3\pi}\left(\frac{H_fa_f^2}{H_Na_N^2}\right)^2
		}^{\mbox{surface term}}
	\, \, \, \mbox{for }  a_{N} \ll  a_f \le a_{eq} \mbox{ and for }  m_r=1 \mbox{, }  
	\nonumber \\ &&
	D_I(a_f) \approx
	-\frac{\alpha_1 w + \alpha_2}{\textcolor {red}{8}} 
	\left(\ln{\frac{a_f}{a_N}}\right)^2
		\overbrace{
			+\frac{1}{12\pi}
			\left[
			3
			\left(\ln{\frac{a_f}{a_N}}\right)^2
			- 6
			\ln{\frac{a_f}{a_N}}
			+2-2\alpha_4
			\right]
		}^{\mbox{surface term}}
	\nonumber \\ &&
	\, \, \, \mbox{for }  a_{N} \ll  a_f \le a_{eq} \mbox{ and for }  m_r=2 \mbox{, }  
	\nonumber \\ &&
	D_I(a_f) \approx
	-( {\alpha_1 w + \alpha_2}) 
	\left[\frac{1}{\textcolor {red}{8}}
	\left({\frac{a_f}{a_N}}\right)^2
	\right]
		\overbrace{
			-\frac{1}{24\pi}
			\frac{a_f}{a_N}
		}^{\mbox{surface term}}
	\nonumber \\ &&
	\, \, \, \mbox{for }  a_{N} \ll  a_f \le a_{eq} \mbox{ and for }  m_r=3 \mbox{, }  
	\nonumber \\ &&
	D_I(a_f) \approx
	-\frac{\alpha_1 w + \alpha_2}{\textcolor {red}{32}} 
	\left({\frac{a_f}{a_N}}\right)^4
		\overbrace{
			-\frac{1}{36\pi}
			\left(\frac{a_f}{a_N}\right)^3
		}^{\mbox{surface term}}
	\, \, \, \mbox{for }  a_{N} \ll  a_f \le a_{eq} \mbox{ and for }  m_r=4 \mbox{, }  
	\nonumber \\ &&
	D_I(a_f) \approx D_I(a_{eq}) + 
	\frac{H_f^2 a_f^4}{ H_0^2 \Omega_r} \left\{ 
	\frac{ H_0^2 \Omega_\Lambda}{H_f^2} \left(\frac{2\pi}{3} \alpha_3 - \frac{ 
		\alpha_2}{12}  \right) \times 
	\right. 
	\nonumber \\ &&
	\left[ 
	\textcolor {red}{\frac{12}{5}} 
	+\textcolor {red}{\frac{12}{13}} \left( \textcolor {red}{\frac{11}{13}} +\ln{\frac{a_f}{a_{eq}}} \right) {\frac{a_{eq}}{a_f}}
	+\textcolor {red}{\frac{12}{13}} \left(\textcolor {red}{\frac{18}{5}+\frac{2}{13}} + \frac{3}{7}\right) \left({\frac{a_{eq}}{a_f}}\right)^{\textcolor {red}{15}/2} 	-\textcolor {red}{\frac{48}{7}}\left({\frac{a_{eq}}{a_f}}\right)^{1/2}
	\right] 
	\nonumber \\ &&
	\left. 
	+ \frac{ 
		\alpha_2}{12} 
	\left[ 
	-\textcolor {red}{\frac{316}{49}}
	-\textcolor {red}{3}\left({\frac{a_{eq}}{a_f}}\right)^{1/2}
	+\textcolor {red}{\frac{6}{7}}\left(\textcolor {red}{\frac{94}{7}}  \textcolor {red}{-10}\ln{\frac{a_f}{a_{eq}}} \right) {\frac{a_{eq}}{a_f}}
	\textcolor {red}{-\frac{287}{9}} \left({\frac{a_{eq}}{a_f}}\right)^{\textcolor {red}{9}/2}
	\right.
	\right.
	\nonumber \\ &&
	\left. 
	\left. 
	\textcolor {red}{-\frac{5}{7}\left({\frac{a_{eq}}{a_f}}\right)^{5}
	+\frac{5}{7}\left({\frac{a_{eq}}{a_f}}\right)^{17/2} }
	\right] \right\} 
	\nonumber \\ &&
		\overbrace{
			+\frac{1}{12\pi}
			\left\{
			\frac{1}{2}\left(\frac{H_fa_f^2}{H_Na_N^2}\right)^2{\frac{a_{eq}}{a_f}}
			+\frac{3}{2}
			\left[ 
			c_1^2 - 
			{\frac{a_{eq}}{a_f}}
			\left(
			c_1^2 +4 c_1c_2
			\left(
			\sqrt{\frac{a_f}{a_{eq}}} -1
			\right)
			+c_2^2 \ln{\frac{a_{eq}}{a_f}}
			\right)
			\right]
			\right.
		}^{\mbox{surface term}}
	\nonumber \\ &&
		\overbrace{
			\left.
			-3 
			\left[ 
			c_1-c_2 \sqrt{\frac{a_{eq}}{a_f}}
			\right]^2
			\right\}
		}^{\mbox{surface term}}
	\nonumber \\ &&
	\, \, \, \mbox{for }  a_{eq} \le  a_f \le a_{\Lambda} \mbox{ and for }  m_r=1\mbox{ and for }  m_m=2 \mbox{. }  
	\nonumber \\ &&
	D_I(a_f) \approx D_I(a_{eq}) + 
	\frac{H_f^2 a_f^4}{ H_0^2 \Omega_r} \left\{ 
	\frac{ H_0^2 \Omega_\Lambda}{H_f^2} \left(\frac{2\pi}{3} \alpha_3 - \frac{ 
		\alpha_2}{12}  \right) \times 
	\right. 
	\nonumber \\ &&
	\left[ 
	\textcolor {red}{-\frac{4}{5}  
	\ln{\frac{a_f}{a_{eq}}} }
	+\textcolor {red}{\frac{188}{75}}  
	-
	\textcolor {red}{\frac{48}{7} }
	\left({\frac{a_{eq}}{a_f}}\right)^{1/2}
	+
\textcolor {red}{\frac{60}{13}} {\frac{a_{eq}}{a_f}}
\textcolor {red}{+\frac{12}{5} \ln{\frac{a_{eq}}{a_N}} \left({\frac{a_{eq}}{a_f}}\right)^{2}}
\textcolor {red}{-\frac{24}{7} \ln{\frac{a_{eq}}{a_N}} \left({\frac{a_{eq}}{a_f}}\right)^{5/2}}
	\right. \nonumber \\ &&
	\left.
\textcolor {red}{+\frac{12}{13} \ln{\frac{a_{eq}}{a_f}} \left({\frac{a_{eq}}{a_f}}\right)^{3}}
\textcolor {red}{+\frac{3}{5} \left(\ln{\frac{a_{eq}}{a_N}}\right)^2 \left({\frac{a_{eq}}{a_f}}\right)^{4}}
\textcolor {red}{+\frac{6}{13} \left(1-\ln{\frac{a_{eq}}{a_N}}\right) \left({\frac{a_{eq}}{a_f}}\right)^{5}}
\textcolor {red}{-\frac{1808}{6825} }
 \left({\frac{a_{eq}}{a_f}}\right)^{\textcolor {red}{15}/2}
	\right.
\nonumber \\ && 
\textcolor {red}{-\frac{12}{13} \ln{\frac{a_{eq}}{a_f}} \left({\frac{a_{eq}}{a_f}}\right)^{17/2}}
\textcolor {red}{+\frac{36}{35} \ln{\frac{a_{eq}}{a_N}} \left({\frac{a_{eq}}{a_f}}\right)^{19/2}}      
	\nonumber \\ &&
	\left. 
\textcolor {red}{+\left[-\frac{6}{13} +\frac{6}{13} \ln{\frac{a_{eq}}{a_N}}-\frac{3}{5} \left(\ln{\frac{a_{eq}}{a_N}}\right)^2\right]\ln{\frac{a_{eq}}{a_N}} \left({\frac{a_{eq}}{a_f}}\right)^{23/2}}      
	\right] 
	\nonumber \\ &&
	+ \frac{ 
		\alpha_2}{12} 
	\left[ 
	-\textcolor {red}{4}
	\textcolor {red}{+8}\left({\frac{a_{eq}}{a_f}}\right)^{1/2}
	+\textcolor {red}{\frac{4}{7}} \left(
	-\textcolor {red}{\frac{243}{7}} 
	- \textcolor {red}{15}\ln{\frac{a_f}{a_{eq}}} \right) {\frac{a_{eq}}{a_f}}
\textcolor {red}{-4 \ln{\frac{a_{eq}}{a_N}} \left({\frac{a_{eq}}{a_f}}\right)^{2}}
\textcolor {red}{+4 \ln{\frac{a_{eq}}{a_N}} \left({\frac{a_{eq}}{a_f}}\right)^{5/2}}
	\right. \nonumber \\ && \left. 
	\left. 
\textcolor {red}{-\frac{60}{7} \ln{\frac{a_{eq}}{a_N}} \left({\frac{a_{eq}}{a_f}}\right)^{3}}
\textcolor {red}{- \left(\ln{\frac{a_{eq}}{a_N}}\right)^2 \left({\frac{a_{eq}}{a_f}}\right)^{4}}
	+
\textcolor {red}{\frac{776}{49}} 
	\left({\frac{a_{eq}}{a_f}}\right)^{\textcolor {red}{9}/2}
\textcolor {red}{-\frac{30}{7} \left(1-\ln{\frac{a_{eq}}{a_N}}\right) \left({\frac{a_{eq}}{a_f}}\right)^{5}}      
	\right.
	\right.
	\nonumber \\ &&
	\left. 
	\left. 
\textcolor {red}{+\frac{60}{7} \ln{\frac{a_{eq}}{a_N}} \left({\frac{a_{eq}}{a_f}}\right)^{13/2}}
\textcolor {red}{+ \left(\left(\ln{\frac{a_{eq}}{a_N}}\right)^2+\frac{30}{7}-\frac{30}{7}\ln{\frac{a_{eq}}{a_N}}\right) \left({\frac{a_{eq}}{a_f}}\right)^{17/2}}
	\right] \right\} 
	\nonumber \\ &&
		\overbrace{
			+\frac{1}{12\pi}
			\left\{
			\frac{3}{2}\left(\frac{H_fa_f^2}{H_Na_N^2}\right)^2{\frac{a_{eq}}{a_f}}
			\left[
			\left(\ln{\frac{a_{eq}}{a_N}}\right)^2
			- 2
			\ln{\frac{a_{eq}}{a_N}}
			+2
			\right]
			\right.
		}^{\mbox{surface term}}
	\nonumber \\ &&
		\overbrace{
			\left.
			+\frac{3}{2}
			\left[ 
			c_1^2 - 
			{\frac{a_{eq}}{a_f}}
			\left(
			c_1^2 +4 c_1c_2
			\left(
			\sqrt{\frac{a_f}{a_{eq}}} -1
			\right)
			+c_2^2 \ln{\frac{a_{eq}}{a_f}}
			\right)
			\right]
			\right\}
		}^{\mbox{surface term}}
	\nonumber \\ &&
	\, \, \, \mbox{for }  a_{eq} \le  a_f \le a_{\Lambda} \mbox{ and for }  m_r=2\mbox{ and for }  m_m=2 \mbox{. }  
	\nonumber \\ &&
	D_I(a_f) \approx D_I(a_{eq}) + 
	\frac{H_f^2 a_f^4}{ H_0^2 \Omega_r (m_r-2)^2} 
	\left({\frac{a_{eq}}{a_N}}\right)^{2m_r-4}
	\left({\frac{a_f}{a_{eq}}}\right)^{2m_m-4} \times 
	\nonumber \\ &&
	\left\{ 
	\frac{ H_0^2 \Omega_\Lambda}{H_f^2} \left(\frac{2\pi}{3} \alpha_3 - \frac{ 
		\alpha_2}{12}  \right)  
	\textcolor {red}{\frac{3}{5}}
	\left[1-\left({\frac{a_{eq}}{a_f}} \right)^{\textcolor {red}{15}/2}
	\right] 
	- \frac{ 
		\alpha_2}{\textcolor {red}{12}} 
	\left[1-\left({\frac{a_{eq}}{a_f}}\right)^{\textcolor {red}{9}/2}
	\right] \right\} 
	\nonumber \\ &&
		\overbrace{
			+\frac{1}{12\pi}
			\left[
			\frac{1}{(m_r-1)^2}
			\frac{H_f^2a_f^4}{H_N^2a_N^4}
			\left({\frac{a_f}{a_{eq}}}\right)^{2m_m-5}
			\left({\frac{a_{eq}}{a_N}}\right)^{2m_r-4}
			\right]
		}^{\mbox{surface term}}
	\nonumber \\ &&
	\, \, \, \mbox{for }  a_{eq} \le  a_f \le a_{\Lambda} \mbox{ and for }  m_r>2\mbox{ and for }  m_m>2 \mbox{. }  
	\nonumber \\ &&
	D_I(a_f) \approx D_I(a_{\Lambda})  
	+ 	\textcolor {blue}{6}
	\frac{H_f^2 a_f^4}{ H_0^2 \Omega_r} 
	  \frac{H_0}{H_f}  
	  \left[\frac{\pi\alpha_3}{3}  
	  \Omega_m 
	  - \frac{		\alpha_2\Omega_m}{12}  \right]
	\left[1-\left({\frac{a_{\Lambda}}{a_f}} \right)^{\textcolor {red}{6}}	\right] 
	\nonumber \\ &&
		\overbrace{
			-\frac{1}{4\pi}
			\left\{
			-\left[\frac{3}{2}\left(\frac{H_0a_0^2}{H_fa_f^2}\right)^2
			\Omega_m {{a_f}} -1
			\right]
			\left[ 
			c_1^2 {\frac{a_{\Lambda}}{a_f}} +
			c_4^2 - 
			{\frac{a_{\Lambda}}{a_f}}
			\left(
			c_4^2 - c_3c_4
			{\frac{a_{\Lambda}}{a_f}} 
			+\frac{1}{3}
			c_3^2 
			\left(
			{\frac{a_{\Lambda}}{a_f}} 
			\right)^2
			\right)
			\right]
			\right.
		}^{\mbox{surface term}}
	\nonumber \\ &&
		\overbrace{
			\left.
			+\left[ 
			c_4-c_3 \left(\frac{a_{\Lambda}}{a_f}\right)^2
			\right]^2 		+\frac{2}{3}
			(	\alpha_4-1)
			\right\}
		}^{\mbox{surface term}}
	\nonumber \\ &&
	\, \, \, \mbox{for }  a_{\Lambda} \le  a_f \le a_{0}=1 \mbox{ and for }  m_r=1\mbox{ and for }  m_m=2  \mbox{, }  
	\nonumber \\ &&
	D_I(a_f) \approx D_I(a_{\Lambda})  
	+ \frac{3}{\textcolor {red}{2}}\frac{H_f^2 a_f^4}{ H_0^2 \Omega_r} 
	\left[\ln{\frac{a_{eq}}{a_N}}+2\right]^2 
\frac{H_0}{H_f}  
	  \left[\frac{\pi\alpha_3}{3}  
	  \Omega_m 
	  - \frac{		\alpha_2\Omega_m}{12}  \right]
	\left[1-\left({\frac{a_{\Lambda}}{a_f}} \right)^{\textcolor {red}{6}}
	\right] 
	\nonumber \\ &&
		\overbrace{
			+\frac{1}{12\pi}
			\left\{
			\frac{3}{2}\left(\frac{H_fa_f^2}{H_Na_N^2}\right)^2{\frac{a_{eq}}{a_f}}
			\left[
			\left(\ln{\frac{a_{eq}}{a_N}}\right)^2
			- 2
			\ln{\frac{a_{eq}}{a_N}}
			\right]
			-3\left[c_4 -c_3\left({\frac{a_{\Lambda}}{a_f}}\right)^2 \right]^2
			\right.
		}^{\mbox{surface term}}
	\nonumber \\ &&
		\overbrace{
			\left.
			+\frac{3}{2}
			\left[ 
			c_4^2 - 
			{\frac{a_{\Lambda}}{a_f}}
			\left(
			c_4^2 -2 c_3c_4
			\left(
			{\frac{a_{\Lambda}}{a_f}} -1
			\right)
			+\frac{c_3^2}{3} 
			\left(
			\left({\frac{a_{\Lambda}}{a_f}}\right)^3
			-1
			\right)
			\right)
			\right]
			-2\alpha_4+2
			\right\}
		}^{\mbox{surface term}}
	\nonumber \\ &&
	\, \, \, \mbox{for }  a_{\Lambda} \le  a_f \le a_{0}=1 \mbox{ and for }  m_r=2\mbox{ and for }  m_m=2  \mbox{, }  
	\nonumber \\ &&
	D_I(a_f) \approx D_I(a_{\Lambda})  
	+ \frac{3}{\textcolor {red}{2}}\frac{H_f^2 a_f^4}{ H_0^2 \Omega_r (m_r-2)^2} 
	\left({\frac{a_{eq}}{a_N}}\right)^{2m_r-4}
	\left({\frac{a_f}{a_{eq}}}\right)^{2m_m-4} \times 
	\nonumber \\ &&
\frac{H_0}{H_f}  
	  \left[\frac{\pi\alpha_3}{3}  
	  \Omega_m 
	  - \frac{		\alpha_2\Omega_m}{12}  \right]
	\left[1-\left({\frac{a_{\Lambda}}{a_f}} \right)^{\textcolor {red}{6}}
	\right] 
	\nonumber \\ &&
		\overbrace{
			+\frac{1}{8\pi}
			\left[
			\frac{1}{m_r-2}
			\frac{H_fa_f^2}{H_Na_N^2}
			\left({\frac{a_f}{a_{eq}}}\right)^{m_m-2}
			\left({\frac{a_{eq}}{a_N}}\right)^{m_r-2}
			\right]^2 \frac{a_{eq}}{a_f}
		}^{\mbox{surface term}}
	\nonumber \\ &&
	\, \, \, \mbox{for }  a_{\Lambda} \le  a_f \le a_{0}=1 \mbox{ and for }  m_r>2\mbox{ and for }  m_m>2  \mbox{, }  
	\label{Rotate.6.C.I.no.inf.5}
\end{eqnarray}
where 
\begin{eqnarray}
	&&
	D_I(a_{eq}) \approx
	\frac{\alpha_1 w + \alpha_2}{\textcolor {red}{18}} 
	\, \, \, \mbox{for }    m_r=1 \mbox{, }  
	\nonumber \\ &&
	D_I(a_{eq}) \approx
	- 	\frac{\alpha_1 w + \alpha_2}{\textcolor {red}{8}} 
	\left(\ln{\frac{a_{eq}}{a_N}}\right)^2
	\, \, \, \mbox{for }   m_r=2 \mbox{, }  
	\nonumber \\ &&
	D_I(a_{eq}) \approx
	- 	\frac{\alpha_1 w + \alpha_2}{\textcolor {red}{8}} 
	\left({\frac{a_{eq}}{a_N}}\right)^2
	\, \, \, \mbox{for }    m_r=3 \mbox{, }  
	\nonumber \\ &&
	D_I(a_{eq}) \approx
	-\frac{\alpha_1 w + \alpha_2}{\textcolor {red}{32}} 
	\left({\frac{a_{eq}}{a_N}}\right)^4
	\, \, \, \mbox{for }    m_r=4 \mbox{, }  
	\label{Rotate.6.C.I.no.inf.6}
\end{eqnarray}
and
\begin{eqnarray}
	&&
	D_I(a_\Lambda) \approx D_I(a_{eq}) + 
	\frac{H_\Lambda^2 a_\Lambda^4}{ H_0^2 \Omega_r} 
	\left\{ 
	\frac{ H_0^2 \Omega_\Lambda}{H_\Lambda^2} \left(\frac{2\pi}{3} \alpha_3 - \frac{ 
		\alpha_2}{12}  \right) 
	\textcolor {red}{\frac{12}{5}} 
	-  \textcolor {red}{\frac{79}{147}} 
		\alpha_2
	\right\} 
	\nonumber \\ &&
	\, \, \, \mbox{for } 
	m_r=1\mbox{ and for }  m_m=2 \mbox{, }  
	\nonumber \\ &&
	D_I(a_\Lambda) \approx D_I(a_{eq}) + 
	\frac{H_\Lambda^2 a_\Lambda^4}{ H_0^2 \Omega_r} 
	\left\{ 
	\frac{ H_0^2 \Omega_\Lambda}{H_\Lambda^2} \left(\frac{2\pi}{3} \alpha_3 - \frac{ 
		\alpha_2}{12}  \right) 
	\textcolor {red}{\left(\frac{188}{75}-\frac{4}{5}\ln{\frac{a_f}{a_{eq}}}\right)} 
	- \frac{ 
		\alpha_2}{\textcolor {red}{3}} 
	\right\} 
	\nonumber \\ &&
	\, \, \, \mbox{for } 
	m_r=2\mbox{ and for }  m_m=2 \mbox{, }  
	\nonumber \\ &&
	D_I(a_\Lambda) \approx D_I(a_{eq}) + 
	\frac{H_\Lambda^2 a_{\Lambda}^4}{ H_0^2 \Omega_r (m_r-2)^2} 
	\left({\frac{a_{eq}}{a_N}}\right)^{2m_r-4}
	\left({\frac{a_\Lambda}{a_{eq}}}\right)^{2m_m-4} \times 
	\nonumber \\ &&
	\left\{ 
	\frac{ H_0^2 \Omega_\Lambda}{H_\Lambda^2} \left(\frac{2\pi}{3} \alpha_3 - \frac{ 
		\alpha_2}{12}  \right)  
	\textcolor {red}{\frac{3}{5}}
	- \frac{ 
		\alpha_2}{\textcolor {red}{12}} 
	\right\} 
	\, \, \, \mbox{for } 
	m_r>2\mbox{ and for }  m_m>2 \mbox{. }  
	\label{Rotate.6.C.I.no.inf.7}
\end{eqnarray}

The parameters $c_1$, $c_2$, $c_3$, and $c_4$ are given by
\begin{equation}
	c_1 =  \sqrt{\frac{f_\theta(a_{eq})}{3}} + c_2 
	\mbox{ , }
	\label{Rotate.c1}
\end{equation}
\begin{equation}
	c_2 =   \frac{\sqrt{2} }{m_m-3/2} 
	\frac{H_f}{H_{eq}} 
	\left(\frac{a_f}{a_{eq}}\right)^{m_m} 
	\mbox{ , }
	\label{Rotate.c2}
\end{equation}
\begin{equation}
	c_3 =  
	\frac{\sqrt{2} }{m_m} 
	\frac{H_f}{H_{\Lambda}} 
	\left(\frac{a_{f}}{a_\Lambda}\right)^{m_m} 
	\mbox{ , }
	\label{Rotate.c3}
\end{equation}
and
\begin{equation}
	c_4 =  c_1   - c_2\left(\frac{a_{eq}}{a_\Lambda}\right)^{m_m-3/2} + c_3
	\mbox{ . }
	\label{Rotate.c4}
\end{equation}

The parameter $\alpha_4$ is a constant of integration on the order of unity.  The calculations used for table \ref{Rotate.T5} take the quantities $\alpha_1 w$, $\alpha_2$, and $\alpha_3$ to be of order unity.

\section*{References}


\begin{thebibliography}{10}

\bibitem{Kramer-Stephani-MacCallum-Herlt:1980}
Dietrich Kramer, Hans Stephani, Malcolm MacCallum, and Eduard Herlt.
\newblock {\em {{Exact} Solutions of Einstein's Field Equations}}.
\newblock VEB Deutscher Verlag der Wissenschaften, Berlin, 1980.

\bibitem{Stephani-Kramer-MacCallum-Hoenselaers-Herlt:2003}
Hans Stephani, Dietrich Kramer, Malcolm MacCallum, Cornelius Hoenselaers, and
  Eduard Herlt.
\newblock {\em {{Exact} Solutions of Einstein's Field Equations}}.
\newblock Cambridge University Press, Cambridge, England, 2nd edition, 2003.

\bibitem{EllisMacCallum69}
George F.~R. Ellis and Malcolm A.~H. MacCallum.
\newblock {``A class of homogeneous cosmological models''}.
\newblock {\em Comm. Math. Phys.}, 12:108--141, 1969.

\bibitem{1996gpst.conf..421K}
V.~A. {Korotky} and Y.~N. {Obukhov}.
\newblock {On Cosmic Rotation}.
\newblock In P.~{Pronin} and G.~{Sardanashvily}, editors, {\em Gravity,
  Particles and Space-time}, pages 421--439, Singapore, 1996. World Scientific.

\bibitem{Chechin:2013}
L.~M. Chechin.
\newblock ``{On} the modern status of the universe rotation problem''.
\newblock {\em Journal of Modern Physics}, {\bf 4}(8A):126--132, 2013.

\bibitem{Ellis-Olive:1983}
John Ellis and Keith~A. Olive.
\newblock ``{Inflatation} can solve the rotation problem''.
\newblock {\em Nature}, {\bf 303}:679--681, 1983.

\bibitem{Barrow-Juszkiewicz-Sonoda:1985}
John~D. Barrow, R.~Juszkiewicz, and D.~H. Sonoda.
\newblock ``{Universal} rotation: how large can it be?''.
\newblock {\em Mon. Not. R. Astron. Soc.}, {\bf 213}:917--943, 1985.

\bibitem{Bayin-Cooperstock:1980}
S.~S. Bayin and F.~I. Cooperstock.
\newblock {``Rotational perturbations of Friedmann universes''}.
\newblock {\em Phys. Rev. D}, 22:2317--2322, 1980.

\bibitem{Collins-Hawking:1973b}
C.~B. Collins and S.~W. Hawking.
\newblock ``{The} rotation and distortion of the universe''.
\newblock {\em Mon. Not. R. Astron. Soc.}, {\bf 162}:307--320, 1973.

\bibitem{Collins-Hawking:1973a}
C.~B. Collins and S.~W. Hawking.
\newblock ``{Why} is the universe isotropic?''.
\newblock {\em The Astrophysical Journal}, {\bf 180}:317--334, 1973.

\bibitem{Ellis:1971}
G.~F.~R. Ellis.
\newblock {``Relativistic cosmology''}.
\newblock In R.~K. Sachs, editor, {\em General relativity and cosmology}, pages
  104--182. Academic Press, New York, 1971.

\bibitem{Ellis:2006}
G.~F.~R. Ellis.
\newblock ``{The} {Bianchi} models: {Then} and now''.
\newblock {\em Gen. Relativ. Gravit.}, {\bf 38}:1003--1015, 2006.

\bibitem{Ellis:2009}
G.~F.~R. Ellis.
\newblock ``{Republication} of: {Relativistic} cosmology''.
\newblock {\em Gen. Relativ. Gravit.}, {\bf 41}:581--660, 2009.

\bibitem{Ellis-Wainwright:1997}
G.~F.~R. Ellis and J.~Wainwright.
\newblock {``Cosmological observations''}.
\newblock In J.~Wainwright and G.~F.~R. Ellis, editors, {\em Dynamical Systems
  in Cosmology}, pages 65--83. The University Press, Cambridge, 1997.

\bibitem{Fennelly:1976}
A.~J. Fennelly.
\newblock ``{Effects} of a rotation of the universe on the number counts of
  radio sources: G{\"o}del's universe''.
\newblock {\em The Astrophysical Journal}, {\bf 207}:693--699, 1976.

\bibitem{Hawking69}
Stephen~W. Hawking.
\newblock {``On the rotation of the universe''}.
\newblock {\em Mon. Not. R. Astron. Soc.}, 142:129--141, 1969.

\bibitem{Jaffe.et.al:2005}
T.~R. Jaffe, A.~J. Banday, H.~K. Eriksen, K.~M. G\'orski, and F.~K. Hansen.
\newblock ``{Evidence} of vorticity and shear at large angular scales in the
  {WMAP} data: A violation of cosmological isotropy?''.
\newblock {\em Astrophysical Journal Letters}, 629:L1--L4, 2005.

\bibitem{Jaffe.et.al:2006}
T.~R. Jaffe, A.~J. Banday, H.~K. Eriksen, K.~M. G\'orski, and F.~K. Hansen.
\newblock ``{Fast} and efficient template fitting of deterministic anisotropic
  cosmological models applied to {WMAP} data''.
\newblock {\em Astrophysical Journal}, 643:616--629, 2006.

\bibitem{Raine-Thomas:1982}
D.~J. Raine and E.~G. Thomas.
\newblock ``{Mach's} principle and the microwave background''.
\newblock {\em Astrophysical Letters}, {\bf 23}:37--45, 1982.

\bibitem{PhysRevLett.117.131302}
Daniela Saadeh, Stephen~M. Feeney, Andrew Pontzen, Hiranya~V. Peiris, and
  Jason~D. McEwen.
\newblock How isotropic is the universe?
\newblock {\em Phys. Rev. Lett.}, 117:131302, Sep 2016.

\bibitem{Su-Chu:2009}
S.-C. Su and M.-C. Chu.
\newblock ``{Is} the universe rotating?''.
\newblock {\em The Astrophysical Journal}, {\bf 703}:354--361, 2009.

\bibitem{Wolfe:1970}
A.~M. Wolfe.
\newblock ``{New} limits on the shear and rotation of the universe from the
  x-ray background''.
\newblock {\em The Astrophysical Journal}, {\bf 159}:L61--L66, 1970.

\bibitem{HAWKING:LUTTRELL:1984}
Stephen~W. Hawking and Julian~C. Luttrell.
\newblock The isotropy of the universe.
\newblock {\em Physics Letters B}, 143(1):83--86, 1984.

\bibitem{Amsterdamski:PhysRevD.31.3073}
Piotr Amsterdamski.
\newblock Wave function of an anisotropic universe.
\newblock {\em Phys. Rev. D}, 31:3073--3078, Jun 1985.

\bibitem{Moss:Wright:PhysRevD.29.1067}
I.~G. Moss and W.~A. Wright.
\newblock Wave function of the inflationary universe.
\newblock {\em Phys. Rev. D}, 29:1067--1075, Mar 1984.

\bibitem{WRIGHT:MOSS:1985:115}
W.A. Wright and I.G. Moss.
\newblock The anisotropy of the universe.
\newblock {\em Physics Letters B}, 154(2):115--119, 1985.

\bibitem{Jones:rotation:problem:2020}
R.~Michael Jones.
\newblock ``{The} rotation problem''.
\newblock {\em General Relativity and Gravitation}, 52(5):1--35, May 2020.

\bibitem{isham1975quantum}
C.J. Isham, R.~Penrose, and D.W. Sciama.
\newblock {\em Quantum gravity: an Oxford symposium}.
\newblock Clarendon Press, Oxford, 1975.

\bibitem{isham1981quantum}
C.J. Isham, R.~Penrose, and D.W. Sciama.
\newblock {\em Quantum gravity 2: a second Oxford symposium}.
\newblock Oxford science publications. Clarendon Press, Oxford, 1981.

\bibitem{smolin:2001:three}
L.~Smolin.
\newblock {\em Three Roads to Quantum Gravity}.
\newblock SCIENCE MASTERS. Basic Books, New York, 2001.

\bibitem{DeWitt:1967a}
Bryce~S. DeWitt.
\newblock Quantum theory of gravity. {I}. the canonical theory.
\newblock {\em Phys. Rev.}, 160:1113--1148, August 1967.

\bibitem{DeWitt:1967b}
Bryce~S. DeWitt.
\newblock Quantum theory of gravity. {II}. the manifestly covariant theory.
\newblock {\em Phys. Rev.}, 162:1195--1239, October 1967.

\bibitem{DeWitt:1967c}
Bryce~S. DeWitt.
\newblock Quantum theory of gravity. {III}. applications of the covariant
  theory.
\newblock {\em Phys. Rev.}, 162:1239--1256, October 1967.

\bibitem{Wheeler68}
John~Archibald Wheeler.
\newblock {``Superspace and the nature of quantum geometrodynamics''}.
\newblock In Cecile~M. DeWitt and John~A. Wheeler, editors, {\em Battelle
  Rencontres, 1967 Lectures in Mathematics and Physics}, pages 242--307. W. A.
  Benjamin, New York, 1968.

\bibitem{Giulini2009}
Domenico Giulini.
\newblock The superspace of geometrodynamics.
\newblock {\em General Relativity and Gravitation}, 41(4):785--815, Apr 2009.

\bibitem{Kiefer2009}
Claus Kiefer.
\newblock Quantum geometrodynamics: whence, whither?
\newblock {\em General Relativity and Gravitation}, 41(4):877--901, Apr 2009.

\bibitem{Kiefer2013}
Claus Kiefer.
\newblock Conceptual problems in quantum gravity and quantum cosmology.
\newblock {\em ISRN Mathematical Physics}, 2013(Article ID 509316):1--17, 2013.

\bibitem{Hartle-Hawking83}
James Hartle and Stephen~W. Hawking.
\newblock {``Wave function of the Universe''}.
\newblock {\em Phys. Rev. D}, 28:2960--2975, 1983.

\bibitem{Halliwell:PhysRevD.38.2468:1988}
Jonathan~J. Halliwell.
\newblock Derivation of the wheeler-dewitt equation from a path integral for
  minisuperspace models.
\newblock {\em Phys. Rev. D}, 38:2468--2481, Oct 1988.

\bibitem{PhysRevD.96.106005}
Justin~C. Feng and Richard~A. Matzner.
\newblock From path integrals to the wheeler-dewitt equation: Time evolution in
  spacetimes with a spatial boundary.
\newblock {\em Phys. Rev. D}, 96:106005, Nov 2017.

\bibitem{Kiefer:1991}
Claus Kiefer.
\newblock {``On the meaning of path integrals in quantum cosmology''}.
\newblock {\em Annals of Physics}, 207:53--70, 1991.

\bibitem{Halliwell-Hartle:1990}
Jonathon~J. Halliwell and James~B. Hartle.
\newblock {``Integration contours for the no-boundary wave function of the
  universe''}.
\newblock {\em Phys. Rev. D}, 41:1815--1834, 1990.

\bibitem{Friedmann:1922}
Alexander Friedmann.
\newblock {\"U}ber die {Kr{\"u}mmung} des {Raumes}.
\newblock {\em Zeitschrift f{\"u}r Physik}, 10:377--386, 1922.

\bibitem{LeMaitre:1927}
Georges {LeMa{\^i}tre}.
\newblock {Un univers homog{\`e}ne de masse constante et de rayon croissant
  rendant compte de la vitesse radiale des n{\'e}buleuses extra-galactiques}.
\newblock {\em Annales de la Soci{\'e}t{\'e} Scientifique de Bruxelles},
  A47:49--59, 1927.

\bibitem{Robertson:1935ApJ}
Howard~Percy {Robertson}.
\newblock {Kinematics and World-Structure}.
\newblock {\em Astrophysical Journal}, 82:284--301, November 1935.

\bibitem{Robertson:1936ApJ-a}
Howard~Percy {Robertson}.
\newblock {Kinematics and World-Structure II.}
\newblock {\em Astrophysical Journal}, 83:187--201, April 1936.

\bibitem{Robertson:1936ApJ-b}
Howard~Percy {Robertson}.
\newblock {Kinematics and World-Structure III.}
\newblock {\em Astrophysical Journal}, 83:257--271, May 1936.

\bibitem{Walker:1937}
Arthur~Geoffrey {Walker}.
\newblock {On Milne's theory of world-structure}.
\newblock {\em Proc. Lond. Math. Soc. Series 2}, 42(1):90--127, 1937.

\bibitem{Ellis-Maartens-MacCallum:2012}
George F.~R. Ellis, Roy Maartens, and Malcolm A.~H. MacCallum.
\newblock {\em Relativistic Cosmology}.
\newblock Cambridge University Press, Cambridge, England, 2012.

\bibitem{Planck:Collaboration:2018:VI:publ}
{Planck Collaboration}.
\newblock Planck 2018 results - {VI.} {Cosmological} parameters.
\newblock {\em A\&A}, 641:A6, 2020.

\bibitem{York72}
James~W. York.
\newblock {``Role of conformal three-geometry in the dynamics of
  gravitation''}.
\newblock {\em Phys. Rev. Lett.}, 28:1082--1085, 1972.

\bibitem{Hawking79}
Stephen~W. Hawking.
\newblock {``The path integral approach to quantum gravity''}.
\newblock In Stephen~W. Hawking and Werner Israel, editors, {\em General
  Relativity, an Einstein Centenary Survey}, pages 746--789. The University
  Press, Cambridge, 1979.

\bibitem{Hartle2005}
James Hartle.
\newblock ``{The} action is infinite for an open cosmology''.
\newblock private communication at the conference, ``Spacetime in action, 100
  years of relativity,'' 31 March 2005, Pavia, Italy, 2005.

\bibitem{Hamilton:2023}
Andrew J.~S. Hamilton.
\newblock ``general relativity, black holes, and cosmology''.
\newblock last viewed 28 November 2023,
  $<$https://jila.colorado.edu/$\sim$ajsh/courses/astr5770\_23/grbook.pdf$>$,
  Dec 2021.

\bibitem{MacCallum-Taub:1972}
M.~A.~H. MacCallum and A.~H. Taub.
\newblock ``{Variational} principles and spatially-homogeneous universes,
  including rotation''.
\newblock {\em Commun. Math. Phys.}, {\bf 25}:173--189, 1972.

\bibitem{Schutz76}
Bernard~F. Schutz, Jr.
\newblock {``Perfect fluids in General Relativity: velocity potentials and a
  variational principle''}.
\newblock {\em Phys. Rev. D}, 2:2762--2773, 1976.

\end{thebibliography}

\end{document}